\documentclass[prb,twocolumn,amssymb,amsmath,superscriptaddress]{revtex4}
\usepackage{graphicx}
\usepackage{dcolumn}
\usepackage{bm}
\usepackage{color} 

\begin{document}
\title{Non-equilibrium dynamics across the BEC-BCS crossover}
\author{G.~Seibold} 
\affiliation{Institut f\"ur Physik, BTU Cottbus-Senftenberg, PBox 101344, 03013 Cottbus,
Germany}
\author{J.~Lorenzana}
\affiliation{Institute for Complex Systems, ISC-CNR, Dipartimento di Fisica,
Universit\'{a} di Roma 'La Sapienza', Piazzale Aldo Moro 5, 00185 Roma, Italy}
 
\begin{abstract}
  We investigate the quench dynamics of strongly coupled superconductors within
  the time-dependent Gutzwiller approximation from the BCS to the BEC regime and  
evaluate the out-of-equilibrium transient  spectral density and optical conductivity relevant for pump probe experiments. 
Fourier transformation of the order parameter dynamics reveals
a frequency $\Omega_J$ which, as in the BCS case, is controlled by
the spectral gap. However, 
we find a crossover from the BCS dynamics to a new strong coupling regime where a characteristic frequency  $\Omega_U$, associated to double occupancy fluctuations controls the order parameter dynamics.  The change of regime occurs close to a dynamical phase transition.
  Both, $\Omega_J$ and $\Omega_U$ give rise to a complex structure of self-driven slow Rabi
  oscillations which are visible in the non-equilibrium optical conductivity where also side bands appear due to the modulation of the double occupancy by  superconducting amplitude oscillations. 
  Analogous results apply to CDW and SDW systems. 
\end{abstract} 

\maketitle

\section{Introduction}
During the past two decades rapid progress has been made in the
study of ultracold fermionic quantum gases, in particular concerning
the realization of a paired BCS state \cite{regal04,ketterle04,bart04}
where the interaction strength can be tuned via Feshbach
resonances.~\cite{chin10}
These systems provide a platform to investigate in a controlled way
the coherent modes of superfluid systems like massive amplitude (``Higgs'') or density modes and Goldstone phase excitations of the order parameter.\cite{behrle} 
Also in condensed matter physics the detection of the superconducting
amplitude mode and charge modes in real time 
has being the  the subject of intense research.~\cite{Mansart2013a,Matsunaga2013,Matsunaga2014}

These experiments have motivated the analysis of the BCS pairing problem
with time-dependent interactions and several 
proposals based on the realization of a suitable
out-of equilibrium dynamics (pump) which is then measured by
a probe pulse.~\cite{kuhn1,kuhn2,manske14,Bunemann2017,collado18,Collado2019,Collado2020,benfatto19}
Within the pseudospin formulation
of Anderson \cite{and58} the problem can be mapped onto an effective
spin Hamiltonian for which the Bloch dynamics can be solved exactly.\cite{bara04,yus05,bara06,alt06,yus06} 

Upon considering the situation with a sudden change
of the pairing interaction (``quench'') the dynamics of the Cooper pair states either dephases or synchronizes.~\cite{bara06} In the dephasing regime which occurs when the (attractive) interaction is reduced or moderately increased, the dynamics is characterized by damped amplitude oscillations for small quenches whereas beyond a critical
quench the pairing amplitude decays to zero. Instead, the synchronization regime occurs  upon increasing the interaction beyond a critical value. In this case a self-sustained dynamical state is reached in which all Cooper pairs states oscillate with the same phase. Off course in the presence of damping the oscillations eventually decay.~\cite{Collado2019}

The BCS pairing problem with a energy (or momentum) independent interaction
corresponds to the weak coupling limit of the attractive Hubbard model
which is used to investigate pairing at larger coupling strength, in particular
the crossover from BCS to BEC, see e.g. Ref.~\onlinecite{randeria14} and
references therein.

In this paper we investigate the spectral properties of the
attractive Hubbard model in non-equilibrium situations
based on the time-dependent Gutzwiller approximation (TDGA).
~\cite{lor1,seibold03,seibold04,seibold08,ugenti10,Markiewicz2010b,fabschi1,fabschi2,bueni13,sandri13,mazza17} In the linear response limit
\cite{lor1,bueni13} this approach gives a very good account
of charge,~\cite{seibold03} magnetic \cite{seibold04} and
pairing \cite{seibold08,ugenti10} fluctuations as compared with exact diagonalization on small clusters.
Also away from linear response\cite{fabschi1,fabschi2,bueni13}
and  despite the
lack of true thermalization the TDGA provides a good description
\cite{sandri13,mazza17} of the order parameter dynamics in the prethermal
regimes and shows good agreement with non-equilibrium dynamical mean-field
approximation.~\cite{eckstein09,werner12,tsuji13,balzer15}

The TDGA has been applied to investigate the dynamics of correlated
paramagnetic states,~\cite{fabschi1,fabschi2,sandri2} order parameter dynamics
of antiferromagnetism,~\cite{sandri13} and superconductivity \cite{mazza12,mazza17}
as a result of an interaction quench from an initial to a final Hubbard interaction ($U_i\rightarrow U_F$).
Interestingly, this approach reveals the occurrence of a dynamical
phase transition at a critical final interaction $U_F=U_c$ which depends on density and on $U_i$. The dynamical transition reflects in 
several features: {\it i)} $U_c$ separates a 'weak' from a 'strong coupling'
regime where the latter is characterized by a decreasing long-time
averaged order parameter for increasing interaction strength whereas
in the weak coupling regime the order parameter follows the quenched interaction
similar to standard time-dependent Hartree-Fock theory;
{\it ii)} At $U_c$ the minimum amplitude of the oscillating
Gutziller renormalization factor approaches zero, thus revealing an underlying
'dynamical localization transition';{\it iii)} At $U_c$ the conjugate phase of
the
double occupancy changes from oscillating around zero to a precession around the
unit circle similar to an estonian swing.  

Here we reveal a further attribute of the TDGA dynamical phase transition,
namely we show that it is characterized by a change of the long-time averaged
spectral gap from a low ($\Omega_J$) to a higher ($\Omega_U$) energy scale.
Here $\Omega_J$ is the characteristic frequency of the pairing correlations
(Gorkov function) while $\Omega_U$ is related to the frequency of the
Gutzwiller double occupancy oscillations. 

We further demonstrate a non-linear mechanism relevant at intermediate and
strong coupling by which oscillations of macroscopic variables  (like the double occupancy) originating from a quench, act back on the superconducting quasiparticles as a periodic drive. 
This produces self-sustained Rabi oscillations
originating from the interplay between $\Omega_J$ and $\Omega_U$ excitations.
Indeed, the  TDGA can be viewed as an  
effective BCS model where the bandwidth is periodically driven by the
macroscopic oscillating variables. In the latter case,  
Rabi oscillations have been demonstrated in Refs.~\onlinecite{collado18,Collado2020}.
  We show how the frequencies $\Omega_J$ and $\Omega_U$
  reveal themselves in the density of states (DOS) and optical conductivity.

Because of the attractive-repulsive transformation\cite{Micnas1990} and the  symmetry of the Hubbard model our  results for superconductivity  at half-filling apply also to spin and charge density wave states.  

The paper is organized as follows. In Sec. \ref{sec2} we give
a brief derivation of the TDGA based on a time-dependent
variational principle. Sec. \ref{sec3} is devoted to the analysis
of the quench dynamics, also in comparison to the weak-coupling
BCS limit, and we discuss the structure of the time-averaged DOS in
Sec. \ref{sec4}. The appearance of self-sustained Rabi oscillations
is demonstrated in Sec. \ref{sec5} while in Sec. \ref{sec6} we show
how these excitations reflect in the optical conductivity.
We conclude our discussion in Sec. \ref{sec6}.

\section{Formalism}\label{sec2}
We study the attractive Hubbard model
\begin{equation}
  H=\sum_{k,\sigma}\varepsilon_k c_{k,\sigma}^\dagger c_{k,\sigma}
  +U\sum_r n_{r,\uparrow}n_{r,\downarrow}\label{eq:hi}
\end{equation}
where electrons with dispersion $\varepsilon_k$ on a lattice (number of sites $N$)
interact via a local attraction $U<0$. We are interested in the dynamics after a quench in the interaction. 

\subsection{Equations of Motion}
The  evolution is obtained  variationally by means of the time-dependent  Gutzwiller wave-function $$|\Psi_G\rangle=\hat P_G|BCS \rangle, $$ with $\hat P_G$ and $|BCS\rangle$ the time-dependent Gutzwiller projector and BCS state. The
variational solution of the time-dependent Schr\"odinger equation
can be obtained by requiring the action $S=\int dt L$ to be stationary
with the following real Lagrangian \cite{bueni13}
\begin{equation}\label{eq:lagr}
L = \frac{i}{2} \frac{\langle \Psi_G|\dot{\Psi}_G\rangle 
- \langle \dot{\Psi}_G|{\Psi}_G\rangle}{\langle\Psi_G|\Psi_G\rangle}
- \frac{\langle\Psi_G| H |\Psi_G\rangle}{\langle\Psi_G|\Psi_G\rangle}
\end{equation}
which leads to the equations of motion from the standard Euler-Lagrange 
equations.

Equation~(\ref{eq:lagr}) can be evaluated within the Gutzwiller approximation (GA)
\cite{fabschi1,fabschi2,bueni13}
where expectation values of $|\Psi_G\rangle$ can be expressed
as renormalized expectation values in $|BCS \rangle$.  
Superconductivity is then most conveniently incorporated\cite{Medina1991,Sofo1992} by (a) performing
a rotation in charge space to a normal state, (b) applying the Gutzwiller
approximation and (c) rotating the density matrix back to the
original frame (cf. Refs. \onlinecite{goetz108,goetz208}).

The Gutzwiller approximated expectation value of the Hamiltonian in Eq.~(\ref{eq:lagr}) including a chemical potential term is given by, 
\begin{eqnarray}
E^{GA}&=& T_0+T_1-\mu N + U N {{} D}, \label{eq:ega}
 \\
T_0&=&\sum_{k}q_{\parallel}\varepsilon_k\left\lbrack \langle c_{k,\uparrow}^\dagger c_{k,\uparrow}\rangle - \langle c_{-k,\downarrow} c_{-k,\downarrow}^\dagger\rangle +1\right\rbrack, \nonumber \\
T_1&=&\sum_k \varepsilon_k \left\lbrack q_{\perp} \langle c_{k,\uparrow}^\dagger
c_{-k,\downarrow}^\dagger \rangle +q_{\perp}^*\langle c_{-k,\downarrow} c_{k,\uparrow}\rangle \right\rbrack.\label{eq:tsc}
\end{eqnarray}
where $\langle  \rangle $ denotes the  $|BCS \rangle$ expectation value and we defined  the double occupancy,  
\begin{displaymath}
{{} D}=\langle \Psi_G|n_\uparrow n_\downarrow|\Psi_G\rangle\,,
\end{displaymath}
and the regular ($T_0$) and anomalous ($T_1$)  contribution to the kinetic energy. The anomalous contribution is a characteristic of the Gutzwiller approximation or the equivalent slave Boson formulation\cite{Sofo1992} and arises from the rotation in charge space applied to the kinetic term.
The explicit form of the renormalization factors $q_{\parallel}$ and $q_{\perp}$  is given in Appendix~\ref{appendixa}.

The dynamical variables of the problem are the density matrix,
\begin{eqnarray*}
\underline{\underline{R}}(k) &=& \left(
\begin{array}{cc}
\langle c_{k,\uparrow}^\dagger c_{k,\uparrow}\rangle  & \langle c_{k,\uparrow}^\dagger c_{-k,\downarrow}^\dagger\rangle \\
\langle c_{-k,\downarrow}c_{k,\uparrow}\rangle &
\langle c_{-k,\downarrow}c_{-k,\downarrow}^\dagger\rangle
\end{array}\right)\,,
\end{eqnarray*}
the parameter $D$, and its conjugate phase $\eta$ which
vanishes in the GA equilibrium state.

Stationarity of the Lagrangian leads to the following  equations of motion,~\cite{fabschi1,fabschi2,bueni13}
\begin{eqnarray}
  \frac{d}{dt}\underline{\underline{R}}(k)&=&-i\left\lbrack
  \underline{\underline{R}}(k),\underline{\underline{H}}^{GA}(k)\right\rbrack \label{eq:densmat}\\
  \dot{{} D}&=& \frac{1}{N}\frac{\partial E^{GA}}{\partial\eta} \label{eq:dotd}\\
  \dot{\eta}&=& -\frac{1}{N}\frac{\partial E^{GA}}{\partial {} D}\label{eq:doteta}
\end{eqnarray}  
and the Gutzwiller Hamiltonian is evaluated from
\begin{equation}
  \label{eq:hga}
H^{GA}_{nm}(k)=\frac{\partial E^{GA}}{\partial R_{mn}(k)},
\end{equation}
 which  is explicitly shown in Appendix \ref{appendixb}.

Conservation of the energy $E^{GA}(R,{{} D},\eta)$ follows from
\begin{eqnarray*}
  \frac{d E^{GA}}{dt}&=&\sum_{k}\frac{\partial E^{GA}}{\partial R_{nm}(k)}\dot{R}_{nm}(k)
  +\frac{\partial E^{GA}}{\partial {{} D}}\dot{{{} D}}+\frac{\partial E^{GA}}{\partial \eta}\dot{\eta} \\
  &=& -i \sum_{k} Tr\left\lbrace \underline{\underline{H}}^{GA}(k)
    \left\lbrack \underline{\underline{R}}(k), \underline{\underline{H}}^{GA}(k)\right\rbrack \right\rbrace = 0
\end{eqnarray*}  
where the second and third term in the first line cancel because of
Eqs. (\ref{eq:dotd},\ref{eq:doteta}) and the first term vanishes upon
permutating the trace.

\begin{figure}[tb]
\includegraphics[width=8cm,clip=true]{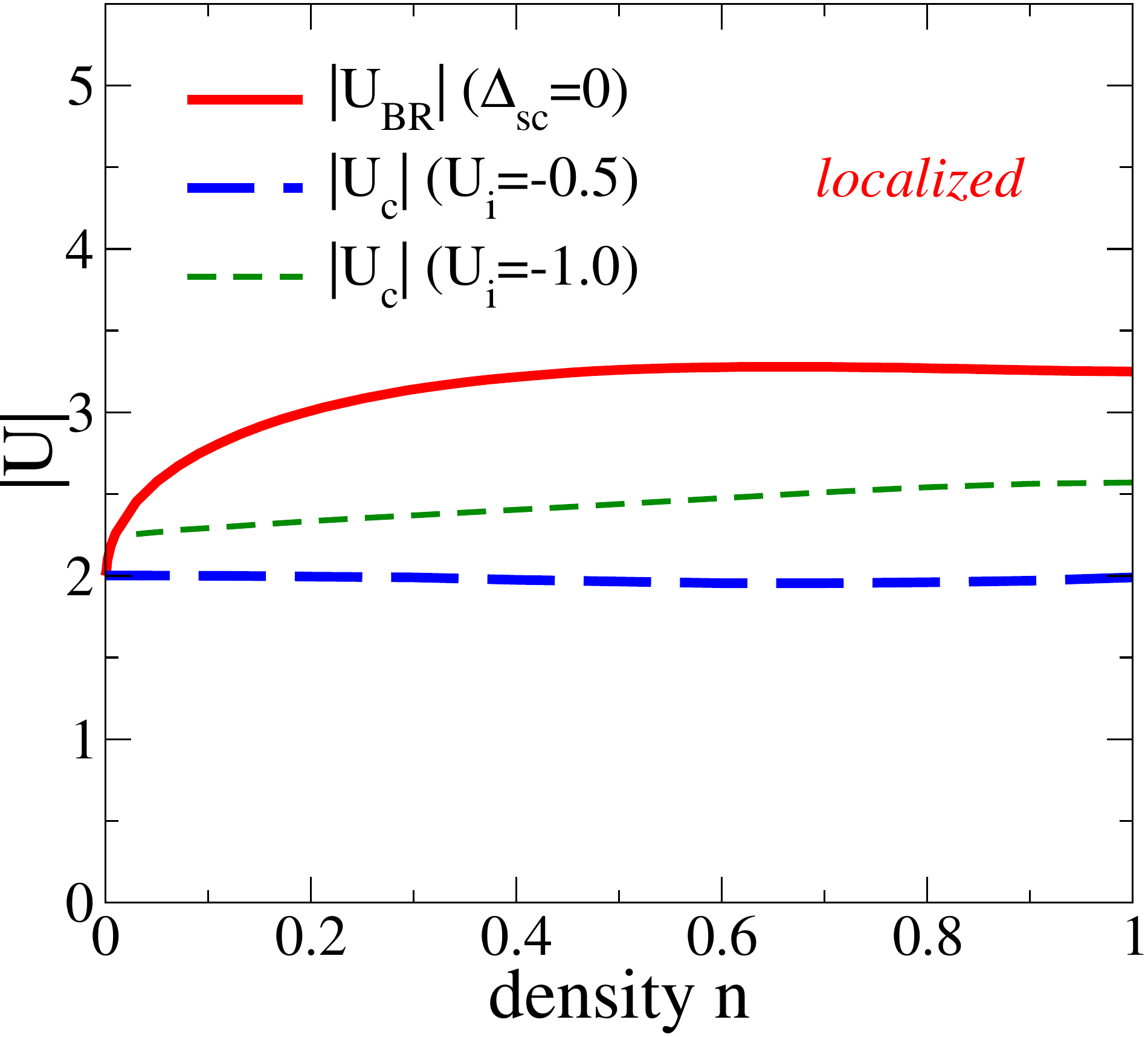}
\caption{Solid red: Brinkman-Rice transition ($q_\parallel=0$) for the attractive, square lattice model in the Gutzwiller approximation restricted to a non-superconducting ground state. Energies are given in units of the half-bandwidth $B=1$. In the superconducting
  system a dynamical phase transition with vanishing $q_\parallel$ occurs
  upon quenching from an initial $U_i$ to a final critical $U_c$.
  The dashed blue and green lines show $U_c$ for $U_i=-0.5$ and
  $U_i=-1$, respectively.}
\label{figphas}                                                   
\end{figure}

It is convenient to  introduce the charge spinor
${\bf J}_k $ with the components
\begin{eqnarray}
 J^x_k &=&  \frac{1}{2}\left(\langle c_{k,\uparrow}^\dagger c_{-k,\downarrow}^\dagger
+c_{-k,\downarrow} c_{k,\uparrow}\rangle\right), \label{eqjx}\\
J^y_k &=& -\frac{i}{2}\left(\langle c_{k,\uparrow}^\dagger c_{-k,\downarrow}^\dagger
-c_{-k,\downarrow} c_{k,\uparrow}\rangle \right), \label{eqjy}\\
J^z_k &=& \frac{1}{2}\left(\langle c_{k,\uparrow}^\dagger c_{k,\uparrow}
+c_{-k,\downarrow}^\dagger c_{-k,\downarrow}\rangle -1\right). \label{eqjz}
\end{eqnarray}
We also define the expectation value of raising and lowering operators, 
\begin{eqnarray}
J^+_k&=&\langle c_{k,\uparrow}^\dagger c_{-k,\downarrow}^\dagger\rangle \\
J^-_k&=& \langle c_{-k,\downarrow} c_{k,\uparrow} \rangle.
\end{eqnarray}
Integrated global quantities will be denoted by dropping the momentum label, i.e. $J^{\mu}\equiv \sum_k  J^{\mu}_k/N$, $J^2\equiv (J^z)^2+(J^x)^2+(J^y)^2$. We will refer to the momentum integrated $J^{\pm}$ as the Gorkov function.

The dynamics of the density matrix can be also expressed via the
dynamics of  Anderson pseudospins in the form of Bloch equations,
\begin{equation}\label{eq:bloch}
  \dot{{\bf J}}_k=2 {\bf b}_k \times {\bf J}_k
\end{equation}
with an effective magnetic field
\begin{equation}\label{eq:bk}
  {\bf b}_k=-(\Delta_k',\Delta_k'',q_{\parallel}(t)\varepsilon_k-\mu).
\end{equation}
Here we defined the real ($\Delta_k'$) and imaginary part ($\Delta_k''$) of the 
spectral gap which is given by the off-diagonal element of the time-dependent Gutzwiller Hamiltonian $\Delta_k(t)\equiv H^{GA}_{12}(k)$ [Eq.~(\ref{eq:h12})].
From Eq.~(\ref{eq:hga}) we see that $\Delta_k(t)$ 
is the conjugate field of the Gorkov function.

In contrast to the BCS case, the gap acquires a momentum
dependence which is determined by the bare dispersion in $H_{12}^{GA}$,
\begin{equation}
  \label{eq:deltak}
\Delta_k\equiv\Delta_\mu+q_{\perp}^*(\varepsilon_k-\mu/q_{\parallel}).  
\end{equation}
 Here we separated  Eq.~(\ref{eq:h12}) into a momentum independent  part $\Delta_\mu$ 
and a momentum dependent part which (unlike a usual momentum-dependent gap)
vanishes at the chemical potential
(see Appendix~\ref{appendixa} for details).

Once the system is taken out of equilibrium both $\Delta_\mu$, $q_{\perp}$ and $q_{\parallel}$ become time dependent. 
In particular,  $q_{\perp}$ is related to fluctuations of the  double occupancy phase $\delta\eta$ [cf. Eq.~(\ref{eq:qper})]. In the weak coupling limit fluctuations
  of the double occupancy phase $\delta\eta$
tend to vanish and one recovers the BCS momentum independent gap since
$q_\perp \sim \delta\eta \to 0$.

The dynamics of the double occupancy ${{} D}(t)$ influences on the
z-component of ${\bf b}_k$ via the renormalization factor 
$q_{\parallel}(t)$ [Eq.~(\ref{eq:qpar})] which will be an essential point in our analysis
of Rabi oscillations in Sec. \ref{sec5}.

\subsection{Static phase diagram}
Before discussing the dynamics we recall the static phase diagram. 
The Gutzwiller approximation for the {\it repulsive}
Hubbard model restricted to a non-magnetic ground state 
yields the well-known Brinkman-Rice transition \cite{br}
at a critical value of $U$ where electrons localize due to the vanishing
of the bandwidth renormalization factor {\em only at half filling}.

In case of the {\it attractive} model restricting to 
a non-superconducting ground state also leads to  a localization transition but now it appears at {\it each} density.~\cite{Medina1991,Sofo1992}
 This is shown in Fig.~\ref{figphas}
where the red line indicates the Brinkman-Rice $U$ above which the ground state is localized. 

The above phase diagram  can be easily understood from the attractive-repulsive transformation\cite{Micnas1990} which maps the negative $U$-Hubbard model into a positive $U$-Hubbard model with a finite magnetization given by $(n-1)/2$.
As it is well known, in the Brinkman-Rice picture a Mott insulator is described as a collection of fully localized spin-1/2 fermions thus effectively neglecting the scale $J$ of magnetic interactions. The Mott states of the negative $U$-Hubbard for arbitrary $n$ can be seen as derivatives of the familiar half-filled positive $U$-Brinkman-Rice insulating state in which a certain number of spins have been flipped to produce a finite magnetization corresponding to  $(n-1)/2\ne 0$.  Thus, for example, 
 a positive $U$-Mott insulating state in which the magnetic configuration is a ferromagnet with a spin-flip ($|\downarrow \downarrow ... \downarrow \uparrow \downarrow... \downarrow\downarrow \rangle $) maps into a 
single composite boson localized 
at the site $i$ of the flipped spin, i.e. the state  $c_{i,\downarrow}^\dagger c_{i,\uparrow}^\dagger|0\rangle$ of the negative $U$-model.
 Clearly, the Mott state reflects the formation of local pairs in the charge language and 
neglecting the magnetic exchange excitations in the positive $U$-language is equivalent to neglecting the boson kinetic energy in the negative $U$-language. Thus, the Brinkman-Rice state corresponds physically to an incoherent state of preformed pairs which would be appropriate  above the critical temperature and below a temperature of the order of $U$ in strong coupling.  Indeed, 
allowing for SC at zero temperature 
the Brinkman-Rice transition is avoided and substituted by the smooth BCS-BEC crossover  
in the stationary state. We anticipate that in a non-equilibrium situation a related dynamical transition appears near the critical $U_F$  depending on the $U_i$.

\section{Quench dynamics}\label{sec3}
In order to study the effect of dimensionality we consider two
systems: (a) A Bethe lattice with infinite coordination number
for which the Gutzwiller approximation becomes exact and
a density of states $\rho(\omega)=\frac{2}{\pi}\sqrt{B^2-\omega^2}$.
(b) A square lattice with nearest-neighbor
hopping which is characterized by a density of
states $\rho(\omega)=\frac{2}{\pi^2 B}{\cal K}(\sqrt{1-\omega^2/B^2})$
and ${\cal K}$ is the complete elliptic integral of the first kind.
All energy scales will be defined with respect to $B\equiv 1$.
In the main part of the paper we will show results for the square lattice
and comment on differences to the dynamics on the Bethe lattice for which
some results are shown in Appendix \ref{appendixc}.

From now on, long-time averages of dynamical quantities $A(t)$ will be denoted by
$\langle A \rangle_T$, i.e. 
\begin{displaymath}
  \langle A \rangle_T=\lim_{t\to\infty}\frac{1}{T}\int_t^{t+T}A(t)
\end{displaymath}
where $T$ comprises a sufficiently large number of oscillations.

\begin{figure}[tb]
\includegraphics[width=8cm,clip=true]{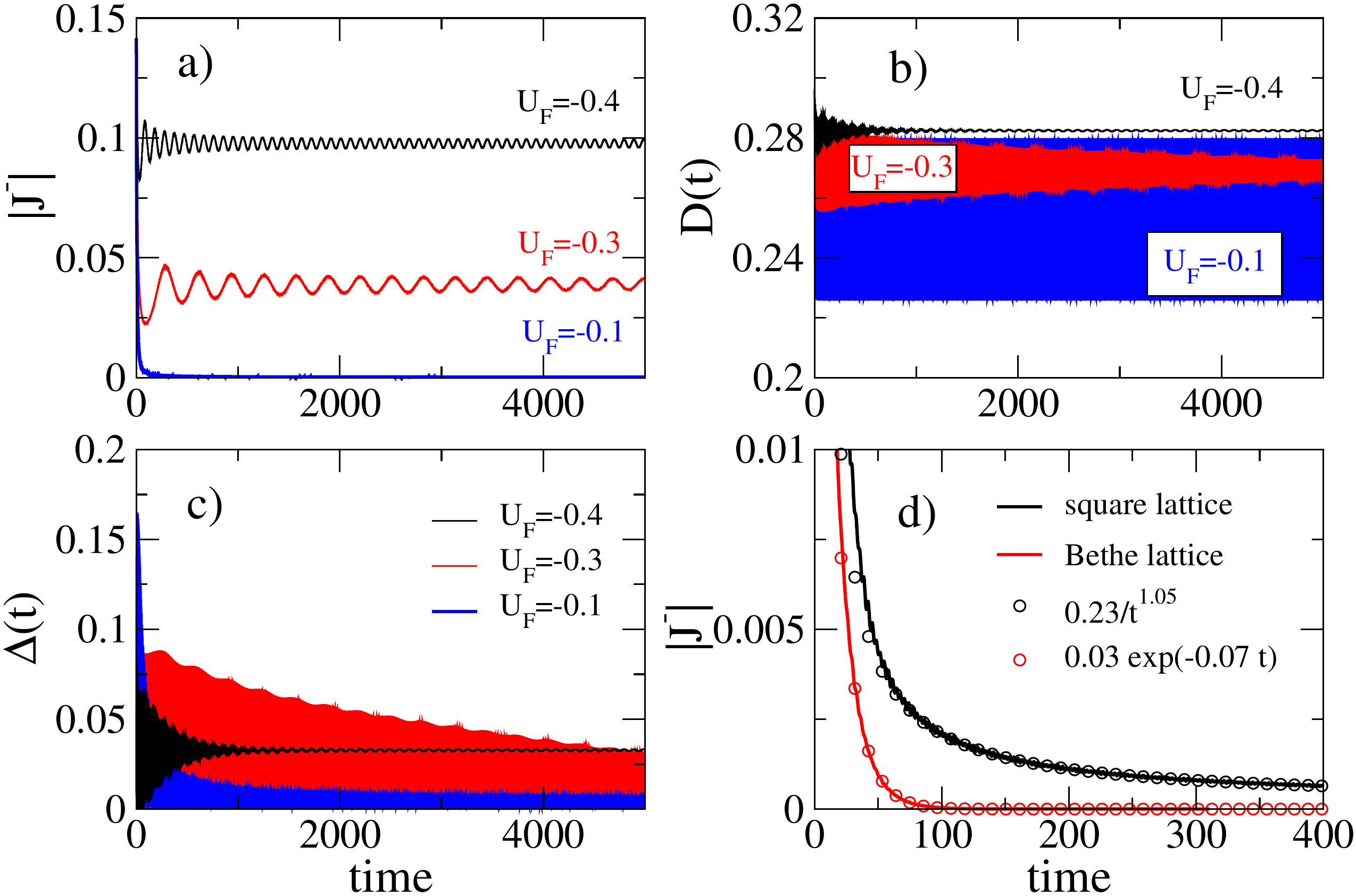}
\caption{Time dependence of the Gorkov function $J^-$ (panels a,d), the
  double occupancy ${{} D}$ (panel b), and the spectral gap parameter at the chemical potential $\Delta_\mu(t)$ (panel c) for a $U$-quench from $U_i=-0.5$ to
  $U_F$ with $|U_F|<|U_i|$. Results in (a-c) are for a half-filled square lattice
  whereas (d) compares the long time behavior of $J^-$ between half-filled Bethe and square lattice for $U_F=-0.1$.  }
\label{fig:1}                                                   
\end{figure}  

\subsection{$|U_F|<|U_i|$ quench}\label{sec:weakcoup}
For small quenches (cf. Fig. \ref{fig:1}a), similar to the
(linearized) BCS dynamics \cite{alt06,yus06}, the Gorkov function
displays a power law, decaying, long-time behavior
\begin{equation}\label{eq:jm}
J^-(t)= J^-_{\infty}\left\lbrack 1+\alpha\cos(2\Delta_\infty t)/\sqrt{\Delta_\infty t}\right\rbrack
\end{equation}
due to dephasing.~\cite{volk74}
We will refer to the dominant frequency of the Gorkov function at long times as
$\Omega_J$. It follows from Eq. (\ref{eq:jm}) that
\begin{equation}\label{eq:2delta}
  \Omega_{J}=2\Delta_\infty\equiv 2 \langle \Delta_\mu(t\to\infty) \rangle_T \,,
\end{equation}
i.e., the frequency of $J^-$ is determined by
the long time limit of the spectral gap $\Delta_\infty$ at the chemical
potential.

Panel (b) of Fig. \ref{fig:1} displays the dynamics of the double occupancy.
Because the frequency is much larger than for the Gorkov function, the main
oscillation is not resolved and only the envelope is visible as the boundary
of the colored regions. We will call the dominant frequency of the double
occupancy $\Omega_U$. 
For the cases in which the Gorkov function oscillates and remains finite at long times (black and red), the dynamics resembles two coupled oscillators with a
fast frequency $\Omega_U$ and a slow frequency $\Omega_J$. Indeed, the slow
frequency of the Gorkov function $\Omega_J$ 
is clearly visible in the envelop of the double occupancy evolution which
shows that $J^-$ and ${{} D}$ are significantly coupled. On the other hand,
since the natural dynamics of $J^-(t)$ is much slower it does not respond
to the fast oscillation of  ${{} D}(t)$ and therefore the fast
oscillations are hardly visible in Fig. \ref{fig:1}a. 
Notice also that the relaxation of ${{} D}(t)$ and $J^-(t)$
occurs on the same time scale.

\begin{figure}[tb]
\includegraphics[width=8.5cm,clip=true]{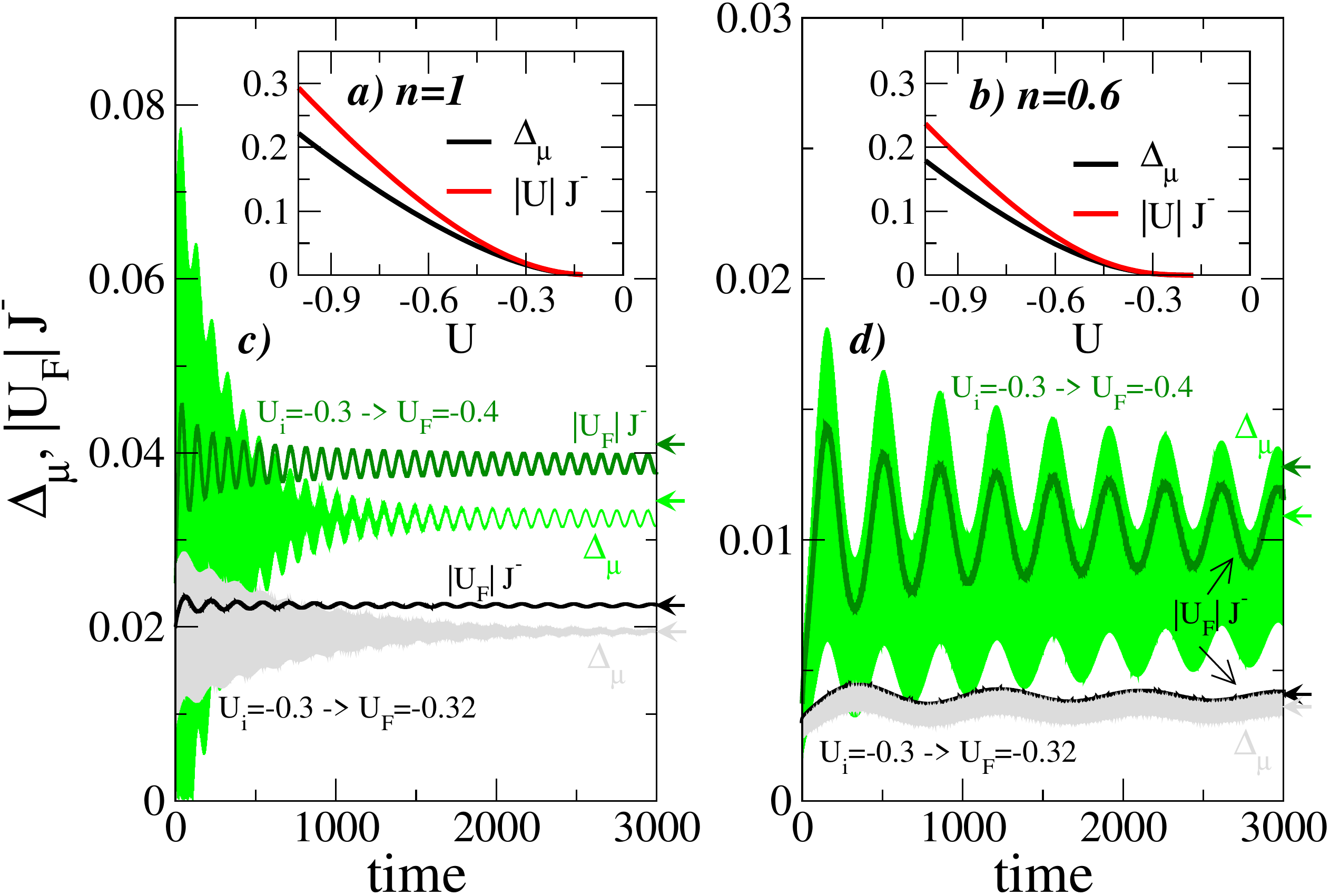}
\caption{Dynamics of the spectral gap  $\Delta_\mu$ and the Gorkov function $J^-$
 near the weak coupling regime. As a reference, the 
inset panels (a,b) compare  the equilibrium
    spectral gap $\Delta_\mu$ (black) with the equilibrium Gorkov function scaled by $|U|$ (red) for densities $n=1$ (a) and $n=0.6$ (b). Panels (c,d):
    Time dependence of $\Delta_\mu$ (light green, grey) and $|U| J^-$ (dark green, black) for a $U$-quench from
    $U_i=-0.3$ to $U_F=-0.4$ and $U_F=-0.32$ for densities $n=1$ (c) and $n=0.6$ (b). Results are for a two-dimensional square lattice. Arrows at the right axis mark the corresponding equilibrium values for $U=U_F$.}
\label{fig:bcs}                                                   
\end{figure}

In case of $J^-(t\to \infty)=0$ (blue) one recovers the situation discussed in
Refs. \onlinecite{fabschi1,fabschi2} where the double occupancy oscillates between the two extrema ${{} D}_-$ and ${{} D}_+$ (upper and lower bounds of the blue curve in (b)). 

Panel (d) of Fig. \ref{fig:1} shows the initial stages of the vanishing of 
$J^-$.
For some critical value of $|U_F|<|U_i|$ the Gorkov function dynamically
vanishes and in this limit 
the decay from an initial $J_i^-$ is described by the general
asymptotic behavior derived in Ref. \onlinecite{yus06}
\begin{equation}\label{eq:j-}
  \frac{J^-(t)}{J^-_i}=A(t)e^{-2\alpha J^-_i t}+B(t)e^{-2 J^-_i t}
\end{equation}
where $A(t)$ and $B(t)$ are decaying power laws $\sim 1/t^\nu$ with $1/2 \le \nu \le 2$ and
$0 \le \alpha \le 1$. As shown in the figure, 
the decay in the 2D system follows a $1/t$ law whereas
for the Bethe lattice it is exponential, 
both behaviors being particular cases of Eq.~(\ref{eq:j-}). 

In general in the TDGA and for moderate to large $U_i$, the dynamics of the spectral gap (cf. Fig. \ref{fig:1}c) is determined by both, the fast double occupancy oscillations at frequency $\Omega_U$ (which are not resolved in the figure), and the slower oscillations of the Gorkov function at frequency $\Omega_J$, which are revealed in the envelope of $\Delta(t)$.

\subsection{$|U_F|>|U_i|$ quench}
\subsubsection{Weak and moderate coupling}
\label{sec:textc-moder-coupl}
One way to characterize the weak coupling (BCS) limit of the dynamics is by comparing the spectral gap
$\Delta_\mu$ with the product of the interaction and the Gorkov function 
$J^-$.  For
  small values of the interaction {\it and} small interaction quenches 
one should recover the  BCS dynamics where the two quantities are related by
 $\Delta_\mu=|U| J^-$.  We first check the 
range of validity of this expression at equilibrium in the inset panels of 
Fig.~\ref{fig:bcs}a,b, where both sides of the equation are shown as a function of the interaction. We see that in this case this relation holds when both quantities become exponentially small i.e. for small attractive interaction.

In order to study the crossover in the non-equilibrium case we show in Fig.~\ref{fig:bcs}c,d the dynamics of $\Delta_\mu$ and $|U| J^-$
  for interaction quenches from $U_i=-0.3$ to $U_F=-0.4$ and
  $U_F=-0.32$. It can be seen that dynamically at short times 
the difference between 
$\Delta_\mu$ and $|U| J^-$ becomes important, even in a regime where the
equilibrium computation shows small or moderate differences. Indeed, 
$\Delta_\mu$ shows again the fast dynamics due to the double occupancy
fluctuations which are absent in $|U| J^-$.
On the other hand, the asymptotic slow dynamics is similar in both quantities. 

At half-filling (Fig.~\ref{fig:bcs}c)
the $\Delta_\mu$ fast dynamics tends to get damped at large times, so a single frequency dominates the dynamics similar to the case of $|U| J^-$. Away from half-filling (Fig.~\ref{fig:bcs}d) this is not anymore true.

Comparing large ($U_F=-0.4$) and small ($U_F=-0.32$) quenches in Fig.~\ref{fig:bcs} we see that the transient phase extends longer in time for smaller quenches but the associated
    oscillations of the gap decrease in amplitude with a concomitant
    decrease in the difference between $\Delta_\mu$ and $|U| J^-$.
    Away from half-filling the coupling of the gap to the double occupancy
    oscillations is significantly enhanced and the associated fast oscillations
    in $\Delta_\mu$ appear with a much larger decay time (not shown, we find $t\approx 10000$ for
    $U_i=-0.3$, $U_F=-0.4$, and $n=0.6$).
    However, similar to the half-filled case the crossover to the BCS dynamics
    occurs via a decrease of the width of these fast oscillations so that
    the envelope of $\Delta_\mu$ approaches $|U| J^-$ in the limit
    of small interaction quenches (and small $U_i$).
    
In BCS the Larmor precession frequency of pseudospins   
(corresponding to the phase velocity of the momentum resolved Gorkov function  $J^-_k$)  is determined by the $z$ component of the pseudomagnetic field through a Bloch equation as in Eq.~(\ref{eq:bloch}).\cite{yus06,bara06} 
Analogously, here we find that in the regime of Fig.~\ref{fig:bcs} the phase velocity is found to obey $\omega_p =2 (q_\parallel\varepsilon_k-\mu)$ [cf. Eq.~(\ref{eq:bk})].

In general for $U_F<U_i$ and small or moderate interactions  the long-time average values of the Gorkov function is slightly below but close to the equilibrium value as in the BCS case.
This is shown in panels b and d of Figs.~\ref{fig:5} and \ref{fig:6} where the red dots correspond to the $U_i$ values.  Also the long-time average of the double occupancy and the regular kinetic energy, $T_0$, are close to the equilibrium vales (panels a and d). 
Notice that the kinetic energy has also an anomalous part  (insets of Fig.~\ref{fig:7}) which however is much smaller in magnitude.  

Superconducting correlations weakly influence on the
characteristic frequency of the double occupancy oscillations $\Omega_U$
in this regime. In the half-filled system a linear response analysis
\cite{bueni13} yields
$\Omega_U= 4|\varepsilon_0|\sqrt{q_0}$ where $\varepsilon_0$ denotes
the energy (per site) of the non-interacting system and $q_0$ is the
(equilibrium) Gutzwiller renormalization factor.
Fig.~\ref{fig:5gap}a (star symbol) reveals the reasonable agreement of
this estimate with the result of the full calculation (triangles)
for small quenches.

\subsubsection{Strong coupling}

For large quenches $|U_F|/  |U_i|\gg 1$ the BCS dynamics crosses over to a
synchronized regime \cite{bara04,bara06}
characterized by phase locked Cooper pair states and an order parameter
dynamics which oscillates nonharmonically between two extrema $\Delta_-$ and
$\Delta_+$. Remarkably, even in this regime the main frequency of
the order parameter in the pure BCS dynamics 
is determined by the average spectral gap, i.e. it obeys Eq.~(\ref{eq:2delta}).
Although the proof is simple, we are not aware of it in the BCS literature 
so we explicitly show it in Appendix~\ref{appendixb}. The validity of the BCS dynamics requires that $|U_F|,|U_i|\ll 1$. The TDGA approximation allows to relax that restriction and explore the intermediate and large coupling regime.

\begin{figure}[tb]
\includegraphics[width=8.5cm,clip=true]{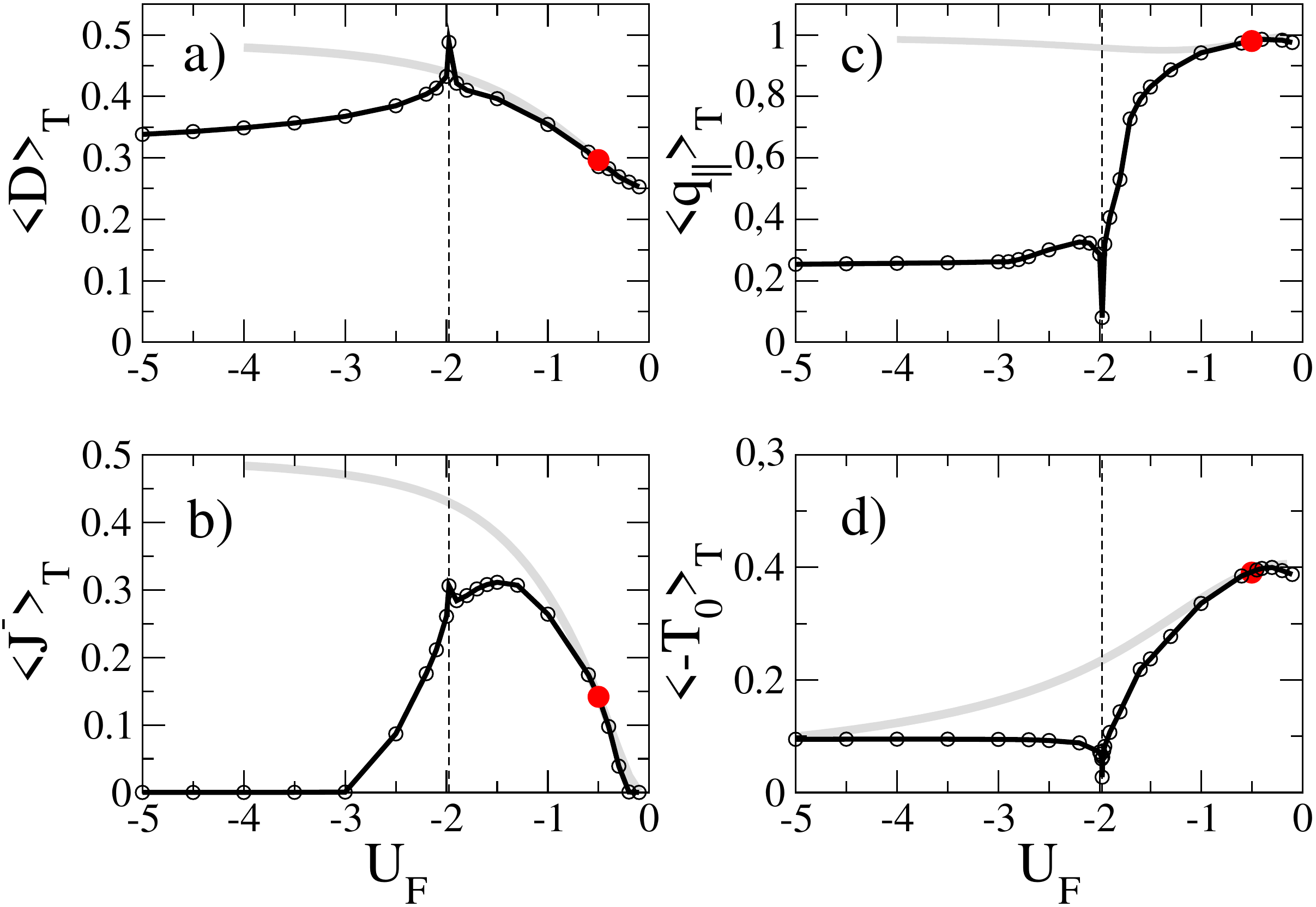}
\caption{Long-time averages of double occupancy (a), Gorkov function (b),
  Gutzwiller renormalization factor (c), regular kinetic energy, $T_0$  (d) for a quench from $U_i=-0.5$ to $U_F$ for a half-filled square lattice.  The red dots correspond to the equilibrium values at $U_i$ and  
  the vertical dotted line indicates the dynamical phase transition at $U_c$.
  The gray lines show the equilibrium value for $U=U_F$.}
\label{fig:5}            
\end{figure}

For large quenches with $|U_F|>|U_i|$ and away from weak coupling
the Gutzwiller dynamics is quite different from the BCS dynamics with the former approaching a dynamical
phase transition at $U_c$.~\cite{fabschi1,fabschi2,sandri13,mazza17}
This is characterized by the dynamics of the phase 
$\eta$ which changes from oscillating around zero to a precession around the
unit circle. Figure~\ref{figphas} compares the density dependence of $U_c$ for two initial $U$ values, $U_i=-0.5$ and $U_i=-1$ with the Brinkman-Rice equilibrium transition. Clearly the typical scale of both transitions is the same.

\begin{figure}[tb]
\includegraphics[width=8.5cm,clip=true]{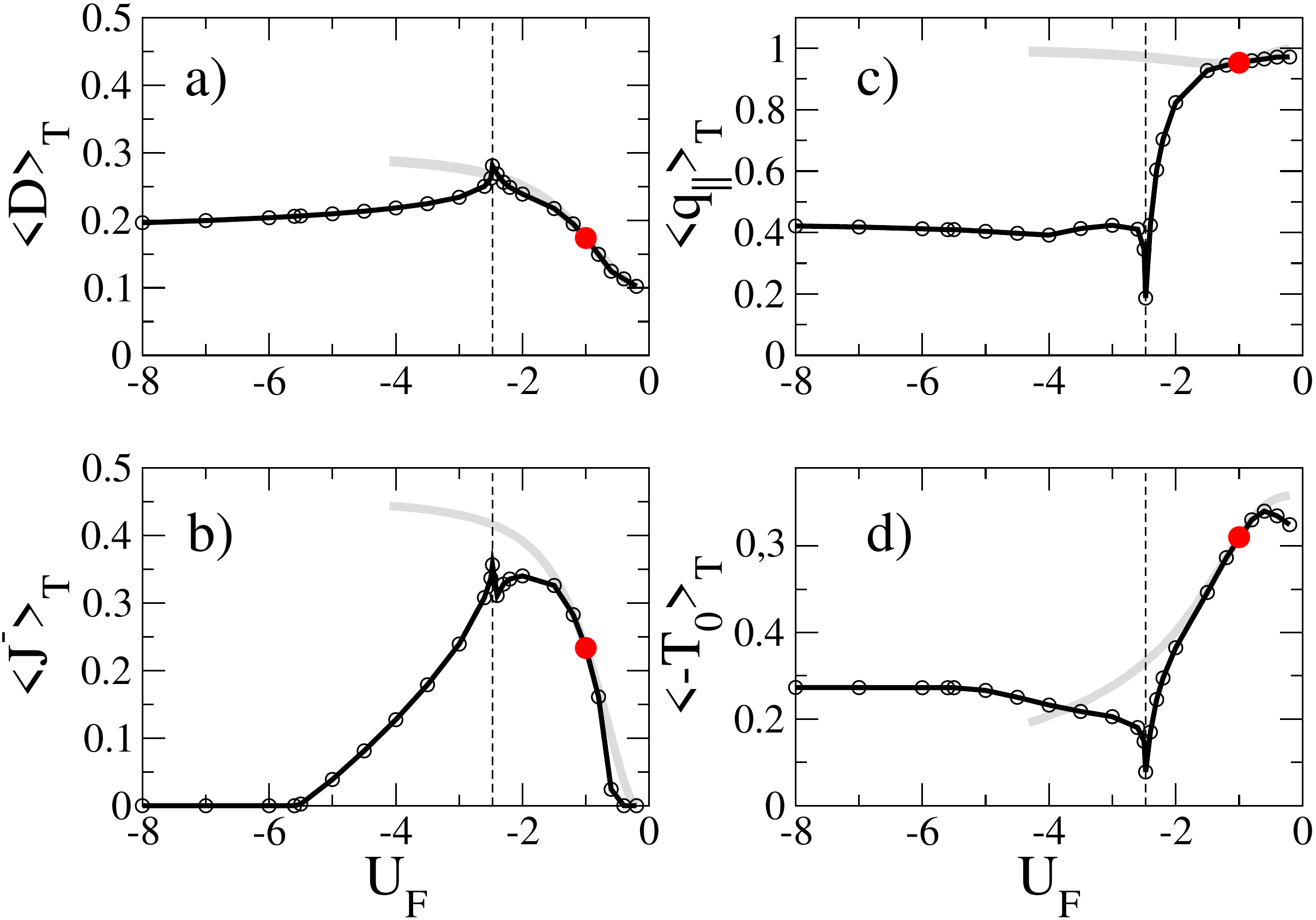}
\caption{Same as Fig. \ref{fig:5} but for concentration $n=0.6$.
}
\label{fig:6}            
\end{figure}

\begin{figure}[tb]
\includegraphics[width=8.5cm,clip=true]{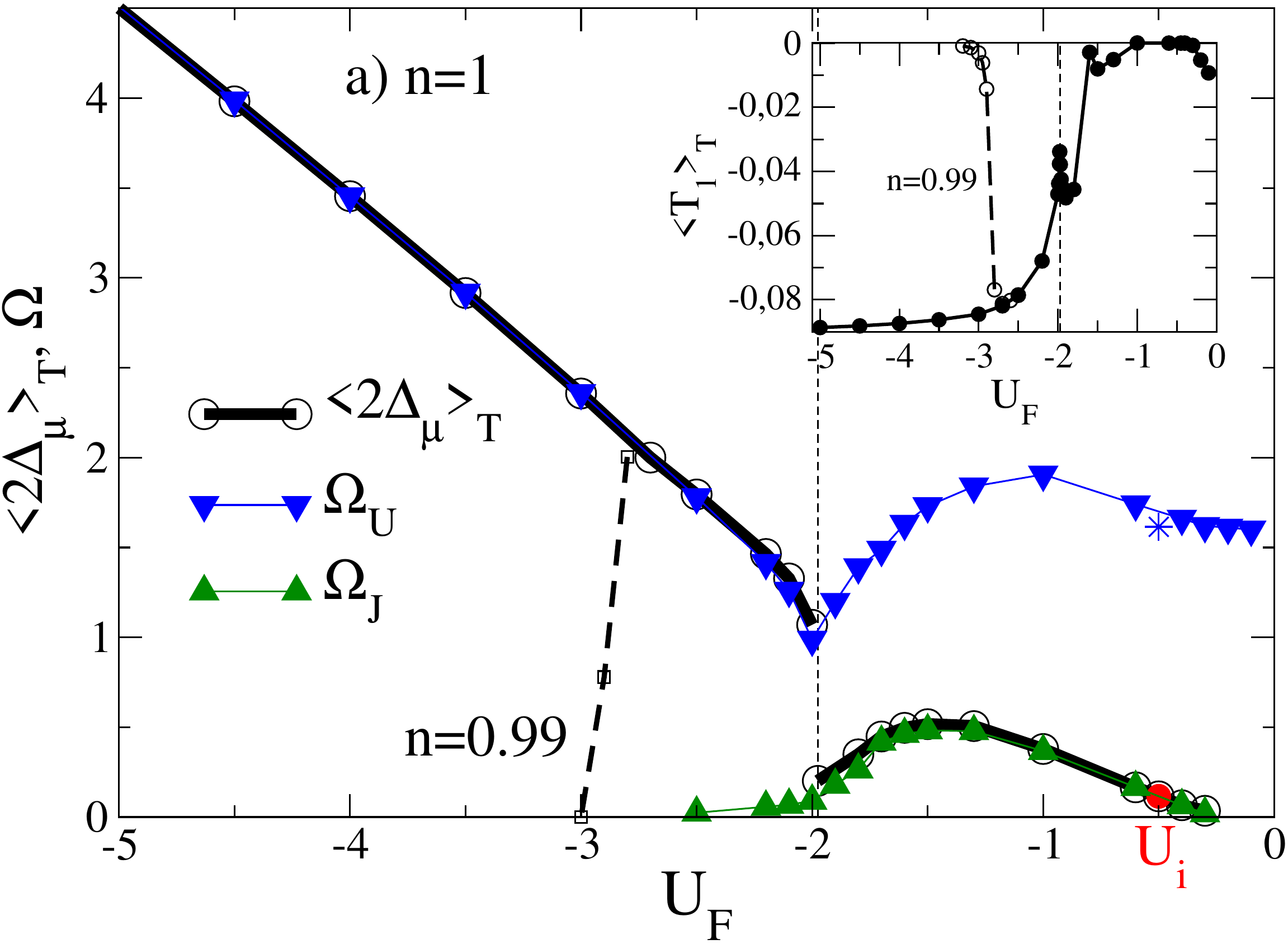}
\includegraphics[width=8.5cm,clip=true]{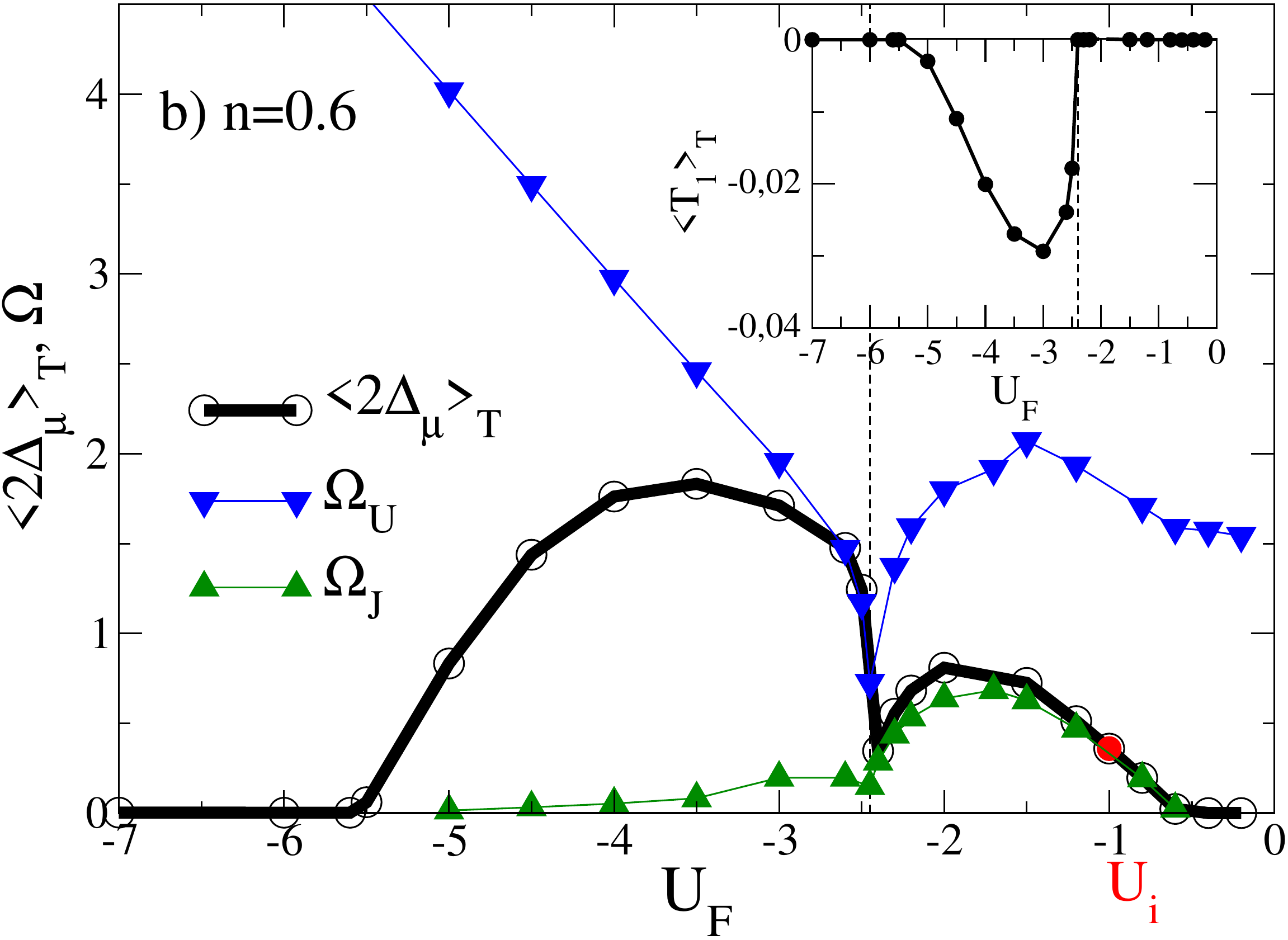}
\caption{
Long-time averages of the spectral gap (black, circles)
  compared to the main frequencies $\Omega_U$ (triangles down, blue)
  and $\Omega_J$ (triangles up, green) in the dynamics
for : (a) quench from $U_i=-0.5$ and $n=1$, 
(b) and $U_i=-1$ and $n=0.6$.  The horizontal axis is the final $U_F$ while the red dots indicate $U_i$.  The dashed line in the inset of panel (a) shows $T_1$ for $n=0.99$. The dashed line in the main panel shows the average of the spectral gap  for $n=0.99$. 
The insets reports the long-time average of the anomalous
kinetic energy  $T_1$ (full line),
cf. Eq. (\ref{eq:ega}).  
The vertical dotted line indicates the dynamical phase transition at $U_c$.
In panel (a) we also show results for $n=0.99$ (dashed line) and the star
symbol indicates the value of $\Omega_U$ from linear response \cite{bueni13}
in the normal system.
}
\label{fig:5gap}            
\end{figure}

Exactly at $U_c$ the average Gutzwiller renormalization factor $q_\parallel$ approaches
zero (cf. Appendix \ref{appendixc} and Fig. \ref{fig:5}c), indicative of an
insulating state. 
This is visible as a maximum in the time averaged double occupancy (cf. Figs. \ref{fig:5}a, \ref{fig:6}a) reaching the value corresponding to full localization ${{} D}=n/2$. At the same time the long-time regular kinetic energy $T_0$, (cf. panel (d) in Figs. \ref{fig:5}, \ref{fig:6} becomes minimum due to the vanishing of $q_\parallel$. Clearly at $U_c$ the system reaches full pairing but quasiparticle dephasing effects completely scramble the kinetic energy of the pairs. 
The spectral gap (Fig. \ref{fig:5gap})
has also a quite interesting behavior. Upon increasing $|U_F|$ it first
follows a BCS like behavior but then it reaches a maximum and starts to
decrease again reaching a minimum at $U_F=U_c$.

As already mentioned in the previous subsection, 
 the long-time average of the Gorkov function (panel (b) of Figs.~\ref{fig:5}, \ref{fig:6})
initially increases with $|U_F|$ and stays slightly below the equilibrium value
for $U\equiv U_F$ (grey line). At $U_c$ the Gorkov function
reaches a local maximum. This may appear paradoxical as it implies that 
the underling BCS state still has a well defined phase and pairing amplitude.
In reality the pairs are fully localized so notwithstanding the phase is well defined, this state is extremely fragile i.e., the cost to scramble it is very
low. More precisely, we will see that the 
 phase stiffness $\rho_s$ tends to vanish. In fact,
  within standard time-dependent BCS theory the non-equilibrium superfluid
  stiffness would be equivalent to
  the average kinetic energy along the direction of the applied vector
  potential.
  Therefore the vanishing of $\langle T_{0}\rangle_T$ at $U_c$ can
  be considered as a ``first indicator'' for the vanishing of 
  $\rho_s$ at $U_c$. However, due to the momentum dependent SC gap the evaluation of $\rho_s$ is more subtle in the TDGA and will be analyzed in Sec. \ref{sec6}.

For larger values of  $|U_F|$  the Gorkov function diminishes and finally vanishes. This suppression of the Gorkov function for large quenches
is opposite to what is obtained within the time-dependent BCS approach
but agrees with non-equilibrium studies within DMFT \cite{werner12}
in the context of quenched antiferromagnetism.
For the half-filled system this vanishing of the Gorkov function
  implies the vanishing of local superconducting correlations. However, isotropic
superconducting s-wave correlations still persist in this regime as can be seen
from the inset to Fig. \ref{fig:5gap} where we report the long-time average
of anomalous kinetic energy correlations $T_1$,
which contribute to the total energy in the TDGA, cf. Eq. (\ref{eq:ega}).
For our two-dimensional system with $\varepsilon_k=-2t[\cos(k_x)+\cos(k_y)]$, 
 Fourier transformation of Eq.(\ref{eq:tsc}) yields a contribution to the energy
which only depends on a symmetric combination of SC correlations between nearest
neighbors, i.e. extended s-wave symmetry, while the bare (i.e. local) s-wave
correlations vanish in the regime of large $U_F$ at half-filling. Moving slightly away from half-filling (dashed line in the inset to Fig. \ref{fig:5gap})
the intersite SC correlations vanish together with the Gorkov function.
In Sec. \ref{sec5} we will analyze this in more detail and show how the double occupancy fluctuations drive the fermions and with increasing strength suppress the average Gorkov function.

\begin{figure}[tb]
\hspace*{-0.5cm}\includegraphics[width=9.5cm,clip=true]{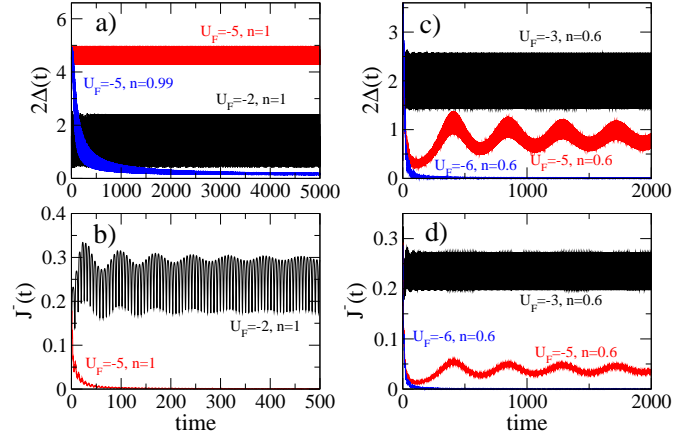}
\caption{ 
Dynamics of the spectral gap $\Delta(t)$ (panels a,c) and
  the Gorkov function $J^-(t)$ (panels b,d) in the regime $|U_F| > |U_i|$
  for $U_i=-0.5$, $n=1$ (panels a,b) and $U_i=-1$, $n=0.6$ (panels c,d).}
\label{fig:7}            
\end{figure}

We now come back to the problem of the relation between 
$2\Delta(t)$ and $J^-$ analyzed in Sec.~\ref{sec:textc-moder-coupl} but now in the 
 strong coupling regime with  $|U_F| > |U_c|$.
 It is apparent from Fig.~\ref{fig:7}a,c that as in the previous cases, 
the dynamics of $2\Delta(t)$ 
is determined by the fast double occupancy oscillations
which are not resolved on the scale of the plot and which give rise to
the filled finite width in the time evolution. 
For the half-filled case (panel a)
the average gap increases with $|U_F|$ (roughly $2\Delta \sim |U_F|$) while the amplitude of the oscillation decreases. 

Figure~\ref{fig:5gap}(a) compares the characteristic frequencies of the
dynamics  $\Omega_U$ (double occupancy, blue triangles)
and $\Omega_J$  (Gorkov function, green triangles).  
Upon increasing $|U_F|$, starting from $|U_i|$,
$\Omega_J$ has a dome shape, somehow similar to the Gorkov  function $\langle J^-\rangle_T$ [Fig.~\ref{fig:5}(b)], until both quantities vanish at $U_F \approx -3$. Instead, $\Omega_U$ remains high and of the order of the bandwidth until for $|U_F|>|U_c|$ it increases linearly with $|U_F|$.
In the same figure we also show the long time average $\langle 2\Delta\rangle_T$
(black lines and circles). For $|U_F|<|U_c|$,  $\langle 2\Delta\rangle_T$  follows $\Omega_J$  (similar to the BCS case) but then it jumps abruptly at 
$U_c$ to $\Omega_U$.  Notice that  
 $\langle 2\Delta\rangle_T=\Omega_U$ holds even in the regime where $\langle J^-\rangle_T =0$. Thus for $|U_F| \gtrsim  3$ the
half-filled system is characterized by finite intersite
  but vanishing local SC correlations and the persistence
  of an average spectral gap which is of the same energy scale than
  the local on-site attraction.

For slight deviations from half-filling and $|U_F| \gg |U_i|$
(cf. dashed line in Fig. \ref{fig:5gap}) the time evolution of the
spectral gap starts from an initial value $\Delta_\mu \sim| U_F|$ but then
relaxes with a $1\sqrt{t}$ behavior to zero ($n=0.99$ in Fig. \ref{fig:7}).

The dynamics of the gap and the Gorkov function for a smaller concentration
$n=0.6$ and $|U_F| > |U_i|$ is shown in panels (c,d) of Fig. \ref{fig:7}
and the corresponding long-time averages in Fig. \ref{fig:6}, \ref{fig:5gap}b.
Similar to the half-filled case the frequency $\Omega_J$ is related
to the spectral gap up to the dynamical phase transition. For
$|U_F|> |U_c|$ the spectral gap initially follows $\Omega_U$ but then goes
through a maximum and vanishes together with the average Gorkov function,
the intersite SC correlations (inset), and $\Omega_J$.

\section{DOS}\label{sec4}
In order to analyze the out-of equilibrium spectral properties we
evaluate the density of states (DOS) obtained
from  the average
\begin{eqnarray}
  \langle \rho(\omega)\rangle_T
  &=&\frac{1}{T}\int_{t_0}^{t_0+T}\!\!\!dt \rho(\omega,t) \label{eq:dos}\\
\rho(\omega,t) &=&
  Im \frac{1}{N\pi}\sum_k\frac{\omega+H_{11}^{GA}(k)}{(\omega-i\eta)^2-(H_{11}^{GA}(k,t))^2-|\Delta_k(t)|^2} \nonumber
\end{eqnarray}
where $t_0$ denotes a time scale after the initial transient dynamics 
and $T$ is 'sufficiently longer' than the characteristic periodicities of the
system. The elements of the Gutzwiller Hamiltonian $H^{GA}$ are defined in
appendix \ref{appendixa}.

\begin{figure}[tb]
\hspace*{-0.5cm}\includegraphics[width=10.5cm,clip=true]{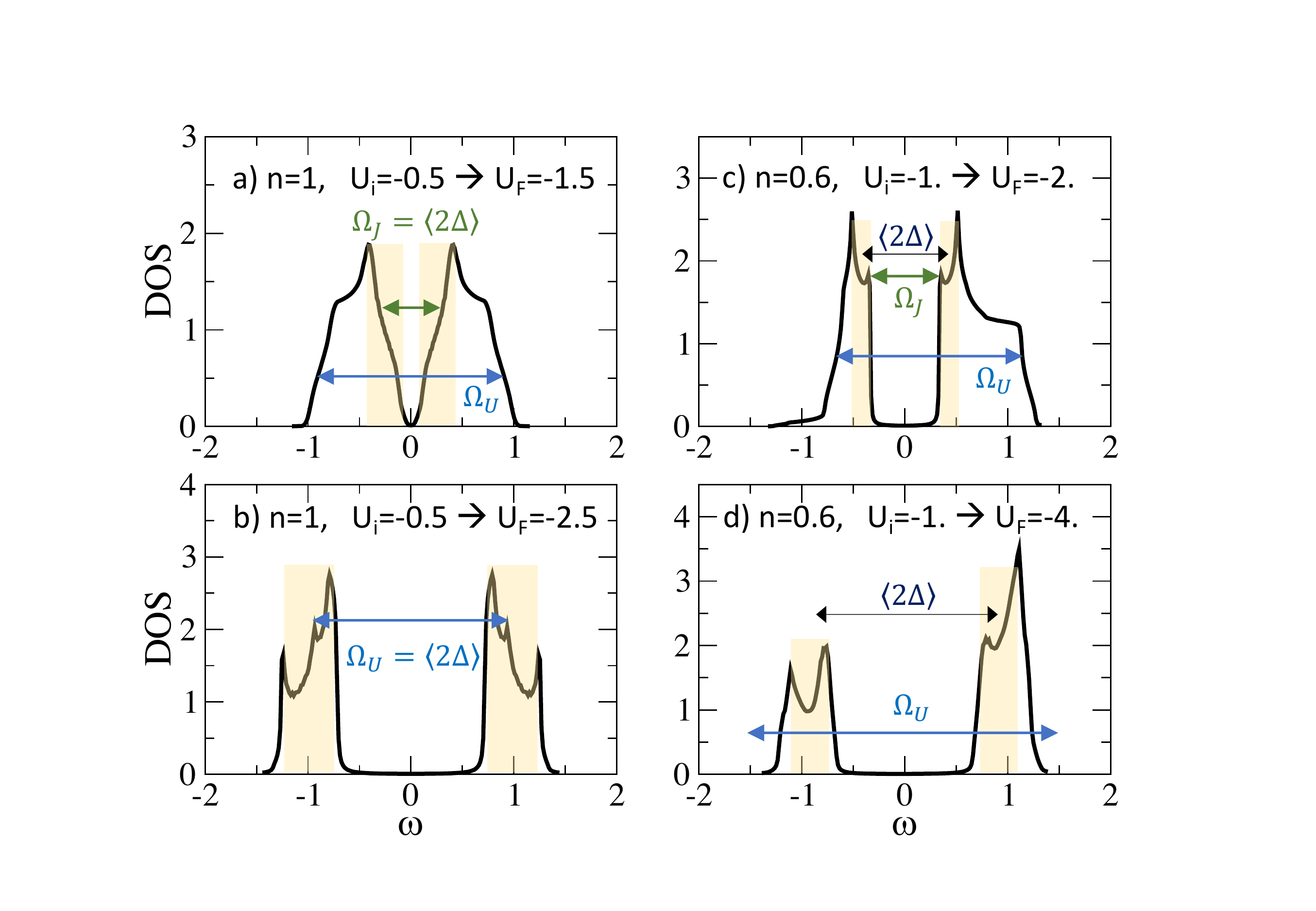}
\caption{Long-time averages of the DOS evaluated from Eq. (\ref{eq:dos})
  for $n=1$ (a,b) and $n=0.6$ (c,d) and different quenches as indicated in
  the panels. Also shown are the characteristic frequencies (where sizeable) $\Omega_J$, $\Omega_U$, cf. Figs. \ref{fig:5}, \ref{fig:6}. The yellow shaded areas indicate the
  variation of the spectral gap in the time-evolution (cf. Fig. \ref{fig:7}).
  Parameters for the evaluation of $\langle \rho(\omega)\rangle_T$, cf.
  Eq. (\ref{eq:dos}): $t_0=500$, $T=100$, $\eta=5\cdot 10^{-3}$.}
\label{fig:dos}            
\end{figure}

Fig. \ref{fig:dos} reports the DOS  for concentrations $n=1$ and $n=0.6$
in case of different quenches $|U_F|>|U_i|$. Clearly the oscillation
amplitude of $\Delta_k(t)$ has a large impact on the low energy
structure of $\langle \rho(\omega)\rangle_T$.
For example, at half-filling and a quench 
$U_i=-0.5 \to U_F=-1.5$ the spectral gap oscillates between $0 \lesssim |\Delta(t)| \lesssim 0.5$ (not shown)
 which gives the impression of a
'd-wave'-shaped gap in the temporal average. Neither $\Omega_J=\langle 2\Delta\rangle_T$ nor $\Omega_U$ are apparent as peculiar feature in the averaged DOS.
On the other, in case $U_i=-0.5 \to U_F=-2.5$ (panel b) 
 the frequency $\Omega_U=\langle 2\Delta\rangle_T$ fits to the transition  between  two peaky structures in the DOS. For even larger values of $|U_F|$ 
 this feature is washed out (not shown).
Note that for the parameters of panel (b) also $\Omega_J\approx 0.024$ is
finite but quite small. 

For the doped system $n=0.6$ and a quench $U_i=-1.0 \to U_F=-2.0$ 
it is the excitation energy $\Omega_J$ which now fits to the transition
between  two peaky structures in the DOS (panel c). For larger quenches 
$\Omega_J$ decreases and does not appear any more in the DOS (panel d).

\section{Frequency mixing and Self-sustained Rabi oscillations}\label{sec5}

\begin{figure}[tb]
\includegraphics[width=9cm,clip=true]{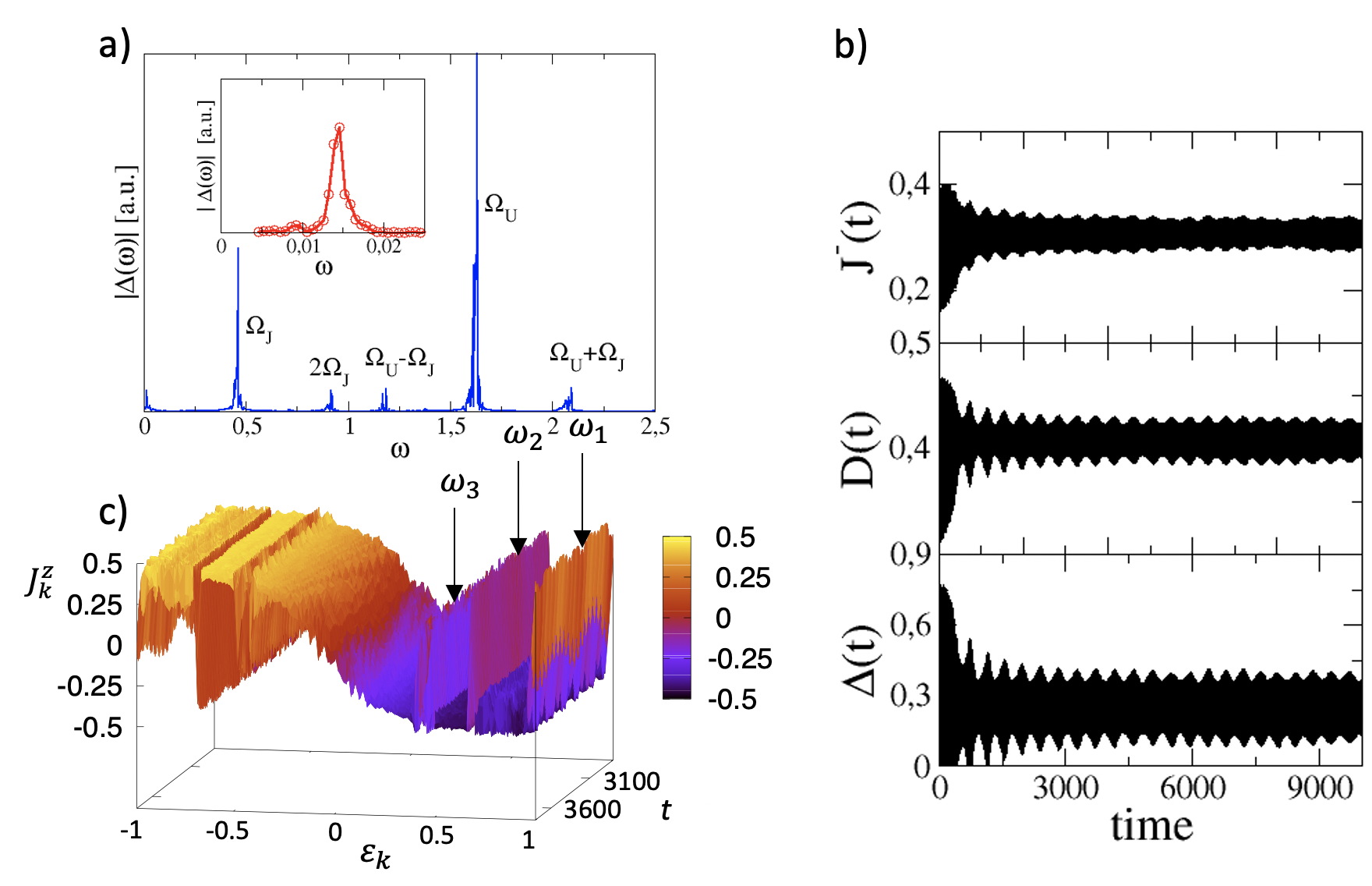}
\caption{a) Fourier spectrum of $\Delta(\omega)$. The inset details the low energy part with the
Rabi excitation.  Quench $U_i=-0.5 \to U_F=-1.6$
  in the half-filled 2D system. b) Energy and time dependence of the pseudospin $J_k^z$ showing population inversion at frequencies $\omega_{1,2,3}$. c) 
From top to bottom: 
Gorkov function, double occupancy, and spectral
  gap.
}
\label{fig:rabi}            
\end{figure}

In BCS the spectral gap after a quench oscillates with its natural frequency $\Omega_J$.  As it is clear from Figs.~\ref{fig:1},\ref{fig:bcs},\ref{fig:7}
the Gutzwiller dynamics is more complex. Besides 
 the frequency $\Omega_J$ the spectral gap responds to the 
 fast oscillations of  the double occupancy with frequency $\Omega_U$. 
Figure~\ref{fig:rabi} shows the Fourier transform of the spectral gap. We see indeed that $\Omega_J$ and  $\Omega_U$ emerge as the prevailing frequencies, but due to the intrinsic non-linearities of the dynamics other frequencies emerge.   

From the equations of motion, we notice that the double occupancy oscillations are seen by the pseudospin degrees of freedom as 'external' drives. 
In fact, the modulation of the bandwidth via $q_\parallel(t)$ [cf. Eq.~(\ref{eq:bk})] adds a time dependence to the effective magnetic field along the z-direction,
$b_k^z=q_\parallel(t)\varepsilon_k$ which we write as $b_k^z=b_k^{z,0}+\delta b_k^z(t)$.
Here $b_k^{z,0}=\langle q_\parallel(t)\rangle_T \varepsilon_k$ is determined
by the temporal average of the renormalization factor and we approximate the time dependent part as $\delta b_k^z(t)\approx \gamma \varepsilon_k \cos(\Omega_D)$ where $\Omega_D$ is the frequency of the drive.
In linear response, the spectral gap responds to fluctuations of the double occupancy, $\delta D$  at the frequency of the driving according to,
\begin{equation}
  \label{eq:lresp}
\delta  \Delta_{\mu}(t)=\chi_{\Delta n} \delta b_k^z(t).  
\end{equation}
where $\chi_{\Delta n}$ is a gap-charge susceptibility (see Ref.~\onlinecite{collado18} for an analogous treatment in the BCS problem).  
In addition, there is an explicit dependence of $\Delta_\mu $ on $D$ through equations Eqs.~(\ref{eq:deltamu}),~(\ref{eq:qpar})-(\ref{eq:Q}). So overall we can write,  
\begin{equation}
  \label{eq:lresp}
\delta  \Delta_{\mu}(t)=\left(\chi_{\Delta n} \frac{\partial b_k^z}{\partial D}+\frac{\partial \Delta_{\mu}}{\partial D}\right)\delta D(t).  
\end{equation}
This explains the appearance of the $\Omega_U$ peak in Fig.~\ref{fig:rabi}(a). 
Extending the expansion to second order in the $\delta D(t)$ and   $\delta J^{\pm}(t)$  fluctuations explains the 
$2\Omega_J$ and the $\Omega_U\pm \Omega_J$ peaks. In fact,
Raman like matrix elements $\partial\chi_{\Delta n}/{\partial J^{\pm}}$ produce Stokes and anti-Stokes responses at $\Omega_U\pm\Omega_J $ and the second harmonic frequency $2\Omega_J$ is generated from $\delta J^{+}(t)\delta J^{-}(t)$ terms which are already present in $\delta b_k^z(t)$ through Eq.~(\ref{eq:qpar}).

Besides these linear and Raman like processes 
another slower characteristic frequency appears 
when one examines the dynamics in very long time windows.  
For example, for quenches $|U_i| < |U_F| < |U_c|$ in the regime where
$\langle J^-\rangle_T$ and $\Omega_J$ are maximum ($U_F \sim -1.6$ in Fig.~\ref{fig:5})   one observes very slow oscillations in
the envelope of all dynamical quantities as shown in Fig.~\ref{fig:rabi}b
for the half-filled system and a quench $U_i=-0.5 \to U_F=-1.6$.
This new frequency is not directly related to the previous ones. Indeed, 
the Fourier transform of this oscillation yields $\Omega_{slow}\approx 0.014$
whereas the frequencies of Gorkov function and double occupancy are
$\Omega_J=0.46$, $\Omega_U=1.63$, see Fig. \ref{fig:rabi}c.
The slow frequency seems to decrease upon approaching $U_c$.

In order to shed some light on this excitation we show in Fig.~\ref{fig:rabi}b the pseudospin dynamics of $J_k^z$ as a function of energy.
One observes population inversion at $\omega_1\approx 0.98$, $\omega_2\approx 0.68$,
and $\omega_3 \approx 0.46$.  Such population inversion in the momentum distribution function ($J_k^z$) is characteristic of collective Rabi oscillations occurring in a superconductor subject to a periodic drive.~\cite{collado18,Collado2020} In the case of a pure BCS dynamics 
as considered in Ref.~\cite{collado18} and for a band width drive 
collective Rabi oscillations are due to states at 'resonant' energies
\begin{equation}\label{eq:wr}
  \varepsilon_k\equiv\omega_r=\frac{1}{2}\sqrt{\Omega_D^2-(2\Delta)^2}\,.
\end{equation}
The corresponding pseudospin will then perform a precession around $b_k^\perp$, which is
the field component of ${\bf b}_k(t)$
perpendicular to the static (or time-averaged) field ${\bf b}_k^0(t)$.\cite{collado18}
Analogous to magnetic resonance dynamics \cite{slichter} the precession ('Rabi') frequency would then be given by
\begin{equation}\label{eq:rabi}
\Omega_R=\frac{1}{2}b_k^\perp=\gamma\Delta\sqrt{1-(2\Delta/\Omega_D)^2} \,.
\end{equation}

In the present time-dependent Gutzwiller dynamics, drives are generated internally as discussed above.   
Based on these arguments the frequencies $\omega_{1,2,3}$ at which population
inversion is seen in Fig. \ref{fig:rabi}b can be obtained
by replacing in Eq. (\ref{eq:wr}) $\Omega_D$ by combinations of $\Omega_U$ and
$\Omega_J$ and also including the average band width renormalization $\bar{q}=\langle q_\parallel\rangle_T$,
\begin{eqnarray*}
  \omega_1&=&\frac{1}{2\bar{q}}\sqrt{(\Omega_D^{(1)})^2-(2\Delta)^2}\approx 0.98 \hspace*{0.2cm}\mbox{for}\hspace*{0.2cm} \Omega_D^{(1)}=\Omega_U\\
  \omega_2&=&\frac{1}{2\bar{q}}\sqrt{(\Omega_D^{(2)})^2-(2\Delta)^2}\approx 0.67 \hspace*{0.2cm}\mbox{for}\hspace*{0.2cm} \Omega_D^{(2)}=\Omega_U-\Omega_J\\
\omega_3&=&\frac{1}{2\bar{q}}\sqrt{(\Omega_D^{(3)})^2-(2\Delta)^2}\approx 0.49 \hspace*{0.2cm}\mbox{for}\hspace*{0.2cm} \Omega_D^{(3)}=2\Omega_J
\end{eqnarray*}
where for the considered quench we have $\bar{q}\approx 0.79$ and
$2\Delta \approx 0.49 \approx \Omega_J$. The combination $\Omega_U+\Omega_J$ would correspond to a drive outside the available energy spectrum. 

\begin{figure}[tb]
\includegraphics[width=8cm,clip=true]{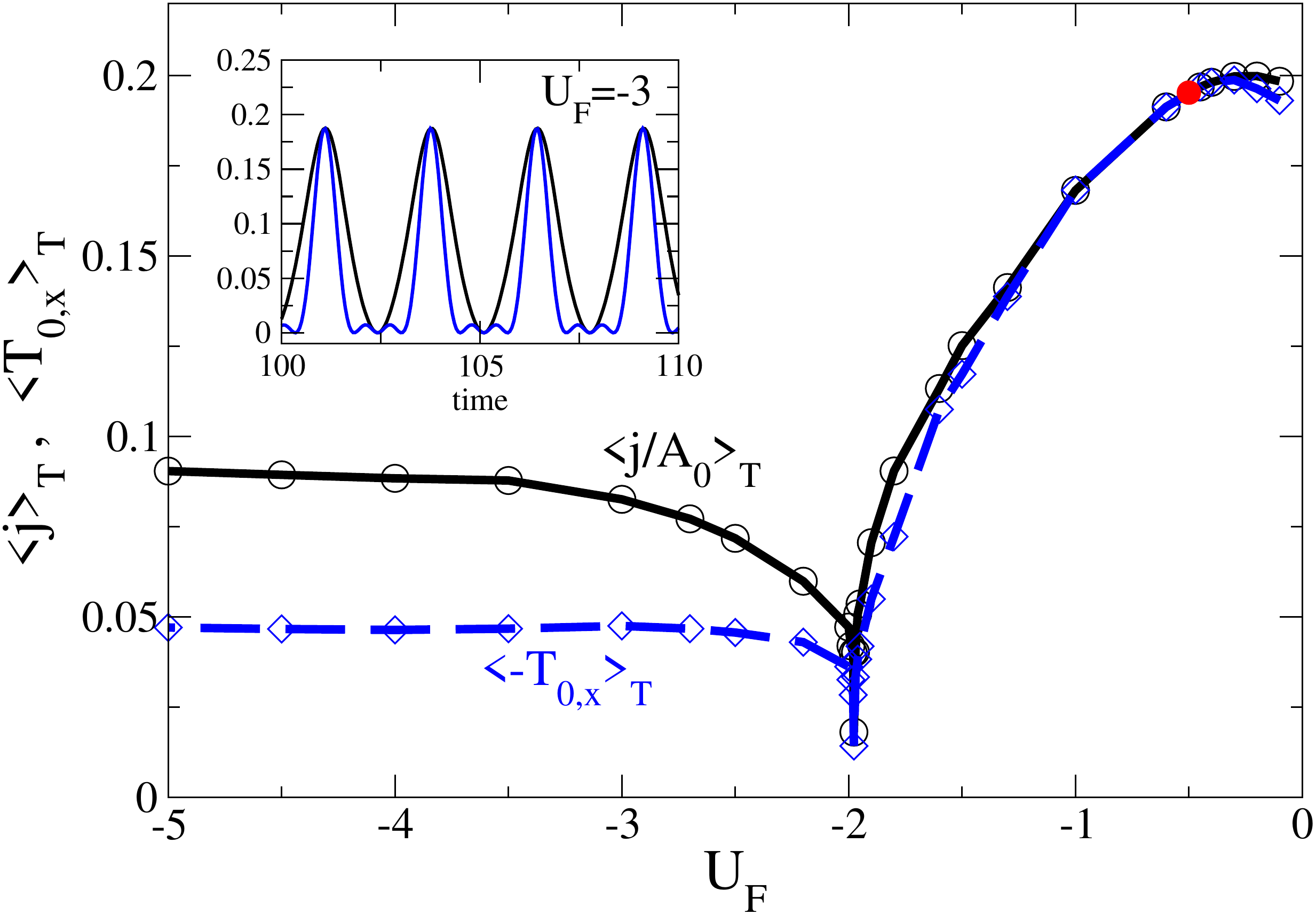}
\caption{Main panel: Superfluid stiffness $\rho_s=j(0)/A_0$ (black, solid) compared to the regular kinetic energy (blue, dashed)
  as a function of the quenched interaction $U_F$ for a half-filled square lattice. The red point indicated the equilibrium value $U_i$.
Inset: Time dependence of both quantities for $U_F=-3$.}
\label{fig:curr}            
\end{figure}

Based on this knowledge we can now ask the
question how these drives are related to the slow Rabi oscillation
visible in Fig. \ref{fig:rabi}.
Generalizing Eq. (\ref{eq:rabi}) to include the bandwidth renormalization
and taking $\gamma\approx 0.05$ as obtained from the width of the renormalization factor
dynamics $q_\parallel(t)$ one obtains
\begin{displaymath}
\Omega^{(1,2,3)}_R=\Delta\frac{\gamma}{\bar{q}}\sqrt{1-(2\Delta/\Omega^{(1,2,3)}_D)^2}
  \approx 0.015, 0.014, 0.013 \,.
\end{displaymath}
Inspection of the low energy Fourier transform of $\Delta(\omega)$ in Fig.~\ref{fig:rabi} reveals a broad excitation centered at $\omega\approx 0.014$ which supports the consistency of our analysis.

\section{Optical conductivity}
\label{sec6}

\begin{figure}[tb]
\includegraphics[width=9cm,clip=true]{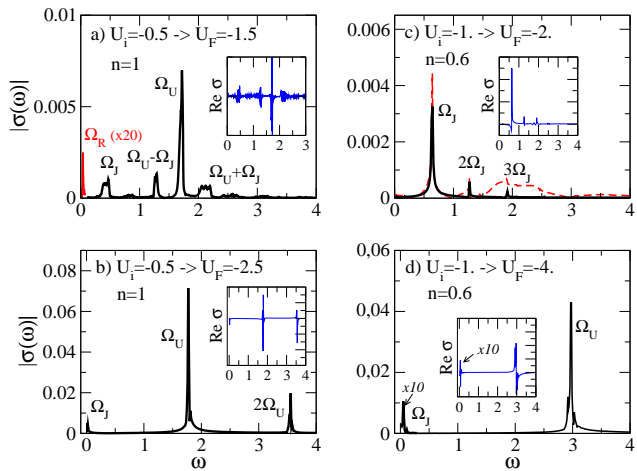}
\caption{Optical conductivity $\sigma(\omega,t_1=50)$
  for the same parameters than in Fig. \ref{fig:dos}. Main panels show
  the magnitude $|\sigma(\omega,t_1)|$ whereas insets report the real part.
  Panel (c) also reports $|\sigma(\omega,t_1)|$ for $t_1=0$ (red dashed)
  which contains the $\Omega_U$ excitations in the transient regime.
  In panel (a) the Fourier transform has been performed for times up
  to $t_{max}=2500$ whereas in panels (b-d) $t_{max}=500$.}
\label{fig:opt}            
\end{figure}

We finally analyze how the characteristic frequencies, discussed in the
previous section, are visible in the
optical conductivity. In the non-equilibrium state
we evaluate this quantity from the current response
\begin{equation}\label{eq:jt}
  j(t)=\int_{t_0}^{t_0+T} dt' \sigma(t,t') E(t')
\end{equation}
to a delta-like electric field $E(t')=A_0\delta(t'-t_1)$ which is applied
within the interval $t_0 < t_1 < t_0+T$ in which the current is measured.
Then the optical conductivity is obtained from the Fourier transformed
of Eq. (\ref{eq:jt}) as \cite{tohyama}
\begin{equation}
j(\omega)=A_0 e^{i\omega t_1 }\sigma(\omega,t_1) \,.
\end{equation}

Within our model the delta-shaped electric field is coupled to the system
by a step-like vector potential
$A(t)=A_0\Theta(t-t_1)$ via the standard Peierls substitution
and the current is evaluated from $j(t)=\delta E^{GA}/\delta A$.
Then the Fourier transform of $j(t)$ is performed for times
$t_1 \lesssim t \lesssim t_1+t_{max}$.

The Peierls substitution for an applied vector
  potential, say along the $x$-direction, induces a shift of momentum
  $k_x \rightarrow k_x + A_x$. Within standard time-dependent BCS theory and
  in the linear response limit this leads to a purely time-dependent
  diamagnetic current which therefore is equivalent to the time-dependent
  kinetic energy. The superfluid stiffness $\rho_s$, defined as the $\omega=0$ component of this current \cite{comment} is then just the time averaged kinetic energy. In the TDGA
  the Peierls substitution also influences on the pairing term
  $\sim q_\perp \varepsilon_k$ [cf. Eqs. (\ref{eq:ega},\ref{eq:tsc}) and appendix \ref{appendixa}] which generates an additional
  pairing component to the current, which is significant in particular at large
  quenches $U_F$ and close to half-filling as can be seen from the inset to Fig. \ref{fig:curr}.
  The main panel of Fig. \ref{fig:curr} compares $\rho_s$ with the regular
  kinetic energy [cf. Eq.~(\ref{eq:tsc})]
  along $x$ as function of the quenched interaction $U_F$. For $U_F\approx U_i$
  both quantities coincide since in this limit the oscillation of the
  double occupancy phase vanish and therefore also $q_\perp \to 0$. Differences
  occur for large quenches and close to half-filling, in particular for $|U_F|>|U_c|$ the normal component of the kinetic
  energy underestimates the stiffness by almost a factor of two for the present
  parameters.

Fig. \ref{fig:opt} shows $\sigma(\omega,t_1)$ for the same
quench situations than analyzed for the DOS, cf. Fig. \ref{fig:dos}.
For a better visualization of the involved frequencies we show
in the main panel the magnitude $|\sigma(\omega,t_1)|$ whereas
the insets display the real part $\sigma'(\omega,t_1)$ with $t_1=50$.
We observe that both frequencies $\Omega_J$ and $\Omega_U$ are visible
in $\sigma(\omega,t_1)$ except for panel c) where the $\Omega_U$
oscillations are already damped for $t<t_1=50$ and are only
visible as a broad feature if the field is switched on already
at $t_1=0$ (red dashed).
A further feature is the coupling of the order parameter to the double
occupancy dynamics, as discussed in the previous section, which is especially
apparent in panel (a) of Fig. \ref{fig:opt} where $\Omega_U$ has two side
peaks at $\Omega_U \pm \Omega_J$ as discussed in the previous section.
This coupling is also present in
panels (b,d) but hardly visible on the scale of the plot due to the
smallness of $\Omega_J$.
In panel (a) we have performed the Fourier transform up to large times
$t_{max}=2500$ which includes several Rabi periodicities. The Rabi
oscillation is visible in $\sigma({\omega})$ though the intensity
is much smaller than those of the main excitations at $\Omega_J$ and $\Omega_U$.
Finally, it should be noted that for large quenches higher harmonics
of $\Omega_U$ appear in the conductivity (cf. panel c).

\begin{figure}[tb]
\includegraphics[width=9cm,clip=true]{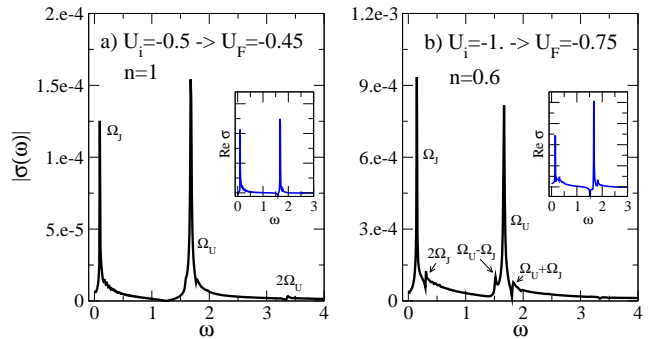}
\caption{Optical conductivity $\sigma(\omega,t_1=50)$
  for quench situations $|U_F|<|U_i|$. Main panels show
  the magnitude $|\sigma(\omega,t_1)|$ whereas insets report the real part.
The Fourier transform has been performed up to $t_{max}=500$.}
\label{fig:optlow}            
\end{figure}  

Similar features can also be seen in Fig. \ref{fig:optlow} which
reports the optical conductivity now for quench situations $|U_F|<|U_i|$.
In both cases the double occupancy oscillations are modulated by
the oscillations of the Gorkov function. For the large quench in panel (b)
the corresponding side bands are clearly visible in $\sigma(\omega)$
together with higher harmonics in $\Omega_J$ and $\Omega_U$.

\section{Conclusions}\label{sec7}
We have analyzed the dynamics of out-of equilibrium superconductivity within
the time-dependent Gutzwiller approximation.
As shown previously \cite{sandri13,mazza17} this approach correctly reproduces
certain aspects of non-equilibrium DMFT
\cite{werner12,tsuji13,balzer15} as the trapping in non-thermal states
and the appearance of two energy scales in the transient dynamics.

In particular, DMFT reveals a sharp crossover in the dynamics of the
Hubbard model upon quenching the non-interacting system to a finite interaction
$U$.~\cite{eckstein09} In the weak coupling regime, below a critical
interaction $U_c$, the double occupancy $D(t)$ relaxes to the almost
thermalized value whereas for strong coupling $D(t)$ recovers and oscillates
with frequency $\sim U$. 

The TDGA captures this feature as a 'dynamical generalization' of the
Brinkman-Rice transition \cite{br} where upon approaching $U_c$ the
period, in which Gutzwiller renormalization factors tend to
zero, logarithmically diverges.~\cite{fabschi1,fabschi2}
In the repulsive Hubbard model the Brinkman-Rice transition is only
present in the half-filled system  
but in the attractive model occurs independent of filling.

Two main frequencies, $\Omega_U$ and $\Omega_J$, determine the dynamical
quantities within the TDGA
which for small quenches are related to the double occupancy and SC pair
correlation dynamics. Here we have shown that the dynamical phase transition
at $|U_c|$ is also associated with a crossover, where the time-averaged SC gap
follows $\Omega_J$ for $|U_F|<|U_c|$ whereas it is bound to $\Omega_U$
in a region
$|U_F|>|U_c|$ which depends on the filling. Interestingly, at half-filling the
average spectral gap keeps following $\Omega_U \sim |U_F|$ for increasing
quenches $|U_F|$, even when the local pair correlations are already suppressed.
We have shown that this regime is instead characterized by intersite SC
correlations (extended s-wave symmetry) which also influence on the
  superconducting stiffness.
It would be interesting to see whether such crossover from local
to extended s-wave superconducting correlations is
also obtained in more exact approaches.

The TDGA can be viewed as a driven BCS model where the drive
acts on the bandwidth via the time dependence of the Gutzwiller renormalization
factors. In an out-of equilibrium situation we have shown that
the characteristic
drive frequency is not only due to the double occupancy dynamics but can be
a linear combination of the basic frequencies $\Omega_U$ and $\Omega_J$.
This yields a consistent explanation for the structure of
low energy Rabi oscillations which can be observed in all dynamical
quantities in certain parameter regimes where the resonant condition
can be fulfilled. Moreover, since for a bandwidth driven BCS model
the increase of the drive amplitude results in a suppression of the
Gorkov function, it is most likely that the same mechanism is also
responsible in the TDGA for the vanishing of $J^{-}$ at large
interaction quenches.

The TDGA does not include thermalization mechanisms so that
in the long-time limit integrated quantities stay either oscillating or
decay due to dephasing, cf. Fig. \ref{fig:7}, whereas in an exact treatment
one expects damping on a time scale $\tau_{th}$. The open question therefore
remains if real systems can be tuned towards a regime where $\tau_{th}$ is
significantly larger than the Rabi periodicity which would allow the
observation of the latter by non-equilibrium spectroscopic methods.

\acknowledgements
G.S. acknowledges financial support from the Deutsche Forschungsgemeinschaft.
J. L. acknowledges insightful discussions with H.P. Ojeda-Collado, C.A. Balseiro and G. Usaj on collective Rabi modes and financial support 
from Italian MAECI through bilateral project AR17MO7, from Italian MIUR through 
Project No. PRIN 2017Z8TS5B, and from Regione Lazio (L. R. 13/08) through project SIMAP.

\appendix
\section{}\label{appendixa}

The renormalization factors in Eq. (\ref{eq:ega}) are given by
\begin{eqnarray}
  q_{\parallel} &=& Q_{+}^2 +\frac{J_{z}^2-J^+J^-}{J^2} Q_{-}^2\label{eq:qpar} \\
  q_{\perp}&=& 2iQ_{-}\frac{J^-}{J}\left\lbrack Q_{+}
 -iQ_{-}\frac{J^z}{J}\right\rbrack\label{eq:qper} 
\end{eqnarray}
with
\begin{eqnarray}
  Q_{+}&=& \sqrt{\frac{\frac{1}{2}-{{} D}+J_z}{\frac{1}{4}-J^2}}
  \left\lbrack \sqrt{{{} D}-J_z-J}\right. \nonumber \\
  &+&\left. \sqrt{{{} D}-J_z+J}\cos\eta \right\rbrack \nonumber\\
&&\\ \label{eq:Q}
  Q_{-}&=& \sqrt{\frac{\frac{1}{2}-{{} D}+J_z}{\frac{1}{4}-J^2}}
  \sqrt{{{} D}-J_z+J}\sin\eta \nonumber\,.
\end{eqnarray}

Then the matrix elements become
\begin{eqnarray}
  H_{11}(k)&=& q_\parallel\varepsilon_k-\mu +\frac{U}{2}(1-\frac{J^z}{J})\\
  &+& \frac{\partial q_\parallel}{\partial J^z}\frac{1}{2N}
\sum_{k'}\varepsilon_{k'} \left\lbrack R_{11}(k')-R_{22}(k')+1\right\rbrack\nonumber \\
&+&\frac{1}{2N}\sum_{k'} \varepsilon_{k'}\left\lbrack
\frac{\partial q_\perp}{\partial J^z}R_{12}({k'})+\frac{\partial q^*_\perp}{\partial J^z}R_{21}({k'})\right\rbrack\nonumber
\\ 
\label{eq:h12}
H_{12}(k)&=& q_\perp^*\varepsilon_k -\frac{U}{2}\frac{J^+}{J}\\
&+&\frac{\partial q_\parallel}{\partial J^-}\frac{1}{N}
\sum_{k'}\varepsilon_{k'} \left\lbrack R_{11}({k'})-R_{22}({k'})+1\right\rbrack\nonumber \\
&+&\frac{1}{N}\sum_{k'} \varepsilon_{k'}\left\lbrack
\frac{\partial q_\perp}{\partial J^-}R_{12}({k'})+\frac{\partial q^*_\perp}{\partial J^-}R_{21}({k'})\right\rbrack\nonumber\\
H_{21}(k)&=& q_\perp\varepsilon_k -\frac{U}{2}\frac{J^-}{J}\\
&+&\frac{\partial q_\parallel}{\partial J^+}\frac{1}{N}
\sum_{k'}\varepsilon_{k'} \left\lbrack R_{11}({k'})-R_{22}({k'})+1\right\rbrack\nonumber \\
&+&\frac{1}{N}\sum_{k'} \varepsilon_{k'}\left\lbrack
\frac{\partial q_\perp}{\partial J^+}R_{12}({k'})+\frac{\partial q^*_\perp}{\partial J^+}R_{21}({k'})\right\rbrack\nonumber\\
H_{22}(k)&=& H_{11}(k).
\end{eqnarray}
The spectral gap in Eq.~(\ref{eq:deltak})
is defined as
$\Delta_k=H^{GA}_{12}(k)=\Delta_\mu+\Delta_k'$ with 
\begin{eqnarray}
  \label{eq:deltamu}
\Delta_\mu&\equiv & \mu \frac{q_\perp^*}{q_\parallel} -\frac{U}{2}\frac{J^+}{J}\\
&+&\frac{\partial q_\parallel}{\partial J^-}\frac{1}{N}
\sum_{k}\varepsilon_{k} \left\lbrack R_{11}({k})-R_{22}({k})+1\right\rbrack\nonumber \\
&+&\frac{1}{N}\sum_{k} \varepsilon_{k}\left\lbrack
\frac{\partial q_\perp}{\partial J^-}R_{12}({k})+\frac{\partial q^*_\perp}{\partial J^-}R_{21}({k})\right\rbrack,\nonumber \\
\Delta_k' &\equiv& \frac{q_\perp^*}{q_\parallel} (q_\parallel\varepsilon_k-\mu).
\end{eqnarray}

\begin{figure}[h!]
\includegraphics[width=8.5cm,clip=true]{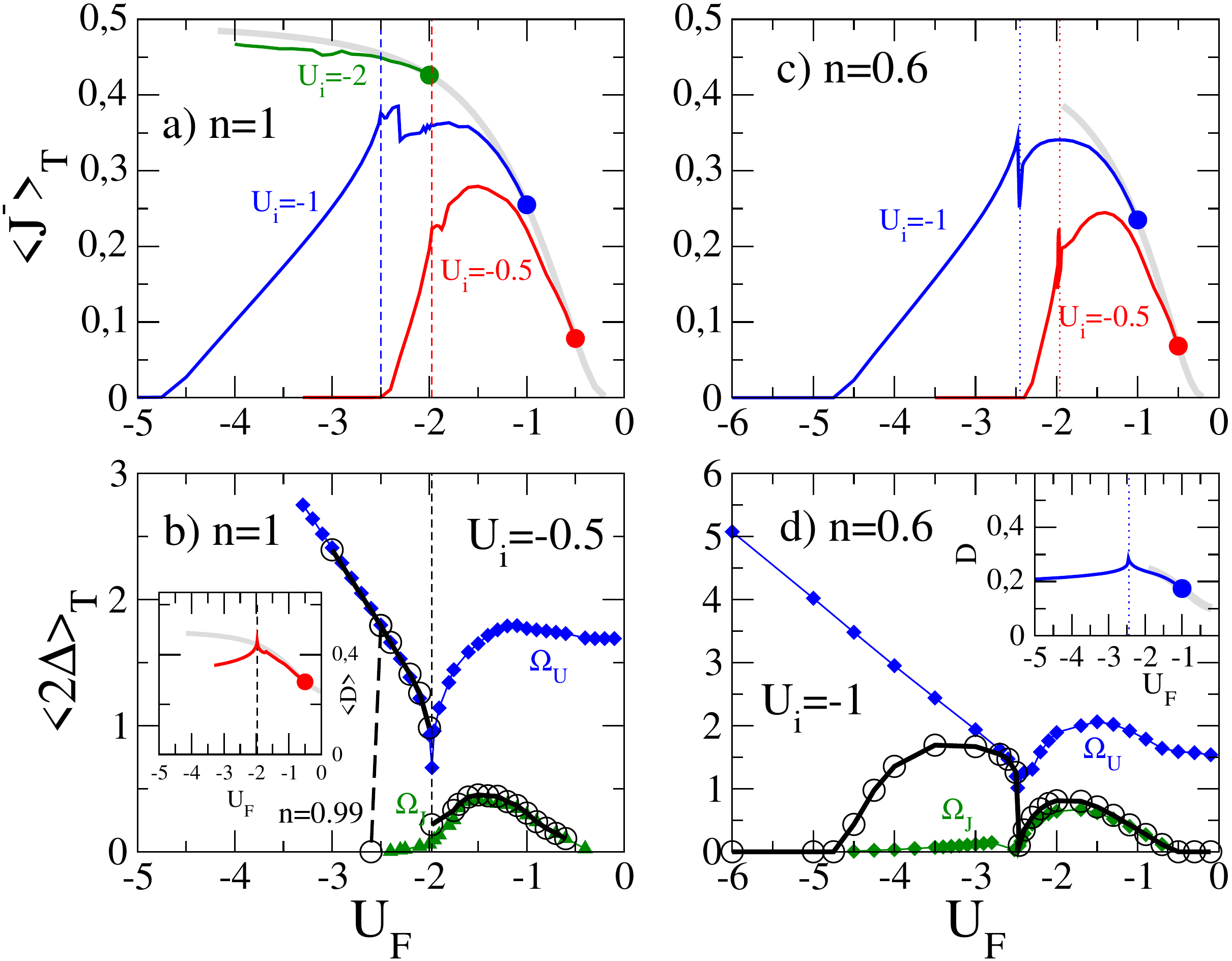}
\caption{Long-time averages of the Gorkov function (a,c) and spectral
  gap (b,d) for the Bethe lattice with infinite coordination number
  and concentrations $n=1$ (a,b) and $n=0.6$ (c,d).
  Panels (b,d) also report the frequencies $\Omega_U$, $\Omega_J$ and the
  insets show the long time average of the double occupancy.}
\label{fig:ap1}            
\end{figure}  

\begin{figure}[h!]
\includegraphics[width=8.5cm,clip=true]{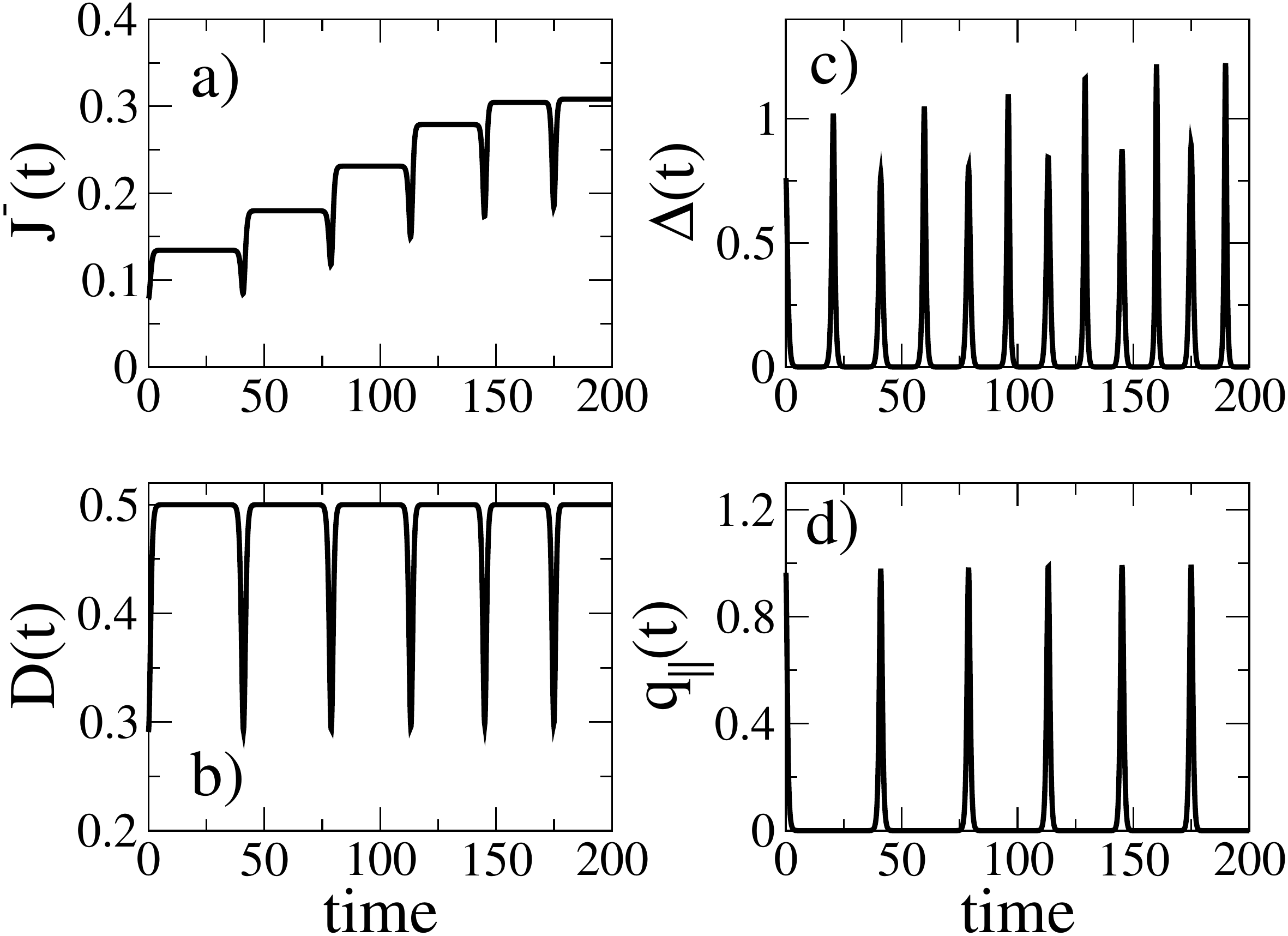}
\caption{Dynamics of the Gorkov function (a), double occupancy
  (b), hopping renormalization $q_\parallel$ (c) and spectral
  gap (d) in close vicinity to $U_c$.
  Results are obtained on a  Bethe lattice with infinite coordination number
  and concentration $n=1$.}
\label{fig:ap2}            
\end{figure}  

\section{}\label{appendixb}
Consider the synchronized regime where the self-consistent BCS dynamics
is governed by the equation \cite{bara04}
\begin{equation}\label{eq:sol}
  \dot{\Delta}^2+\left(\Delta^2-\Delta_-^2\right)\left(\Delta^2-\Delta_+^2\right)=0 \,.
\end{equation}
and shows
soliton solutions of the order parameter oscillating
between $\Delta_- \le \Delta(t) \le \Delta_+$ .
The oscillation period is then determined from
\begin{equation}
  T=2\int_{T_{-}}^{T_{+}}\!\!\!dt = \int_{\Delta_{-}}^{\Delta_+}\frac{d\Delta}{\dot{\Delta}}\,.
\end{equation}
Similarly, the time-averaged order parameter is obtained from
\begin{equation}
  \langle \Delta \rangle_T = \frac{2}{T}\int_{\Delta_-}^{\Delta_+}\!\!d\Delta \, \frac{\Delta}{\dot{\Delta}}
\end{equation}
so that
\begin{equation}
  \langle \Delta \rangle_T = \frac{\omega}{\pi}\int_{\Delta_-}^{\Delta_+}d\Delta\frac{\Delta}{\sqrt{\left(\Delta^2-\Delta_-^2\right)\left(\Delta_+^2-\Delta^2\right)}}=\frac{\omega}{2}
\end{equation}
where we have used Eq. (\ref{eq:sol}) and $\omega=2\pi/T$ is the frequency of the oscillation.
Thus also in the synchronized regime the main frequency $\omega$ of the BCS dynamics 
is determined by the time-averaged spectral gap $2 \langle \Delta \rangle_T$.

\section{}\label{appendixc}
Fig. \ref{fig:ap1} reports the long-time averages of spectral gap
(a,c) and Gorkov function (b,d) for a Bethe-lattice with infinite coordination
number. No qualitative changes occur with regard to the 2-D case
shown in Figs. \ref{fig:5}, \ref{fig:6}.

The time dependence of Gorkov function $J^-(t)$, double occupancy $D(t)$,
hopping renormalization $q_\parallel(t)$, and spectral gap $\Delta(t)$
close to the dynamical phase transition is shown in 
Fig. \ref{fig:ap2}.
The dynamics in this regime is characterized by a
periodic soliton like behavior with 
long localization time periods
where $D(t)$ takes the Brinkman-Rice value ($D=0.5$ for $n=1$) and
the hopping renormalization $q_\parallel$ vanishes.
The time dependence of the phase $\eta(t)$ has a periodicity with twice
the frequency of $D(t)$, $J^-(t)$, and $q_\parallel(t)$ which reflects also
in the dynamics of the  spectral gap $\Delta(t)$ (panel c).


\begin{thebibliography}{cc}
\bibitem{regal04} C. A. Regal, M. Greiner, and D. S. Jin, Phys. Rev. Lett. {\bf 92}, 040403 (2004).
\bibitem{ketterle04} M. W. Zwierlein, C. A. Stan, C. H. Schunck, S. M. F. Raupach, A. J. Kerman, and W. Ketterle, Phys. Rev. Lett. {\bf 92}, 120403 (2004).
\bibitem{bart04} M. Bartenstein, A. Altmeyer, S. Riedl, S. Jochim, C. Chin, J.
  Hecker Denschlag, and R. Grimm, Phys. Rev. Lett. {\bf 92}, 203201 (2004).
\bibitem{chin10} C. Chin, R. Grimm, P. Julienne, and E. Tiesinga, Rev. Mod. Phys. {\bf 82},1225 (2010).
\bibitem{behrle} A. Behrle, T. Harrison, J. Kombe, K. Gao, M. Linkl, J.-S. Bernier, C. Kollath, and M. K\"ohl, Nature Physics 2018, https://doi.org/10.1038/s41567-018-0128-6.

\bibitem{Mansart2013a}
B. Mansart, J. Lorenzana, A. Mann, A. Odeh, M. Scarongella, M. Chergui, and F.
  Carbone, Proc. Natl. Acad. Sci. {\bf 110},  4539  (2013).

\bibitem{Matsunaga2013}
R. Matsunaga, Y.~I. Hamada, K. Makise, Y. Uzawa, H. Terai, Z. Wang, and R.
  Shimano, Phys. Rev. Lett. {\bf 111},  1  (2013).

\bibitem{Matsunaga2014}
R. Matsunaga, N. Tsuji, H. Fujita, A. Sugioka, K. Makise, Y. Uzawa, H. Terai,
  Z. Wang, H. Aoki, and R. Shimano, Science {\bf 345},  1145  (2014).

\bibitem{Bunemann2017}
J. B{\"{u}}nemann and G. Seibold, Phys. Rev. B {\bf 96},  245139  (2017).

\bibitem{Collado2019}
H.~P. {Ojeda Collado}, G. Usaj, J. Lorenzana, and C.~A. Balseiro, Phys. Rev. B
  {\bf 99},  174509  (2019).

\bibitem{Collado2020}
H.~P. {Ojeda Collado}, G. Usaj, J. Lorenzana, and C.~A. Balseiro, Phys. Rev. B
{\bf 101},  054502  (2020).
\bibitem{benfatto19} M. Udina, T. Cea, and L. Benfatto, Phys. Rev. B {\bf 100}, 165131 (2019).
\bibitem{kuhn1} T. Papenkort, V. M. Axt, and T. Kuhn, Phys. Rev. B {\bf 76}, 224522 (2007).  
\bibitem{kuhn2} T. Papenkort, T. Kuhn, and V. M. Axt, Phys. Rev. B {\bf 78}, 132505 (2008).
\bibitem{manske14} H. Krull, D. Manske, G. S. Uhrig, and A. P. Schnyder, Phys. Rev. B {\bf 90}, 014515 2014).
\bibitem{collado18} H. P. O. Collado, J. Lorenzana, G. Usaj, and C. A. Balseiro, Phys. Rev. B {\bf 98}, 214519 (2018).
\bibitem{and58} P. W. Anderson, Phys. Rev. {\bf 112}, 1900 (1958).
\bibitem{bara04} R. A. Barankov, L. S. Levitov, and B. Z. Spivak,  Phys. Rev. Lett. {\bf 93}, 160401 (2004).
\bibitem{yus05}  E. A. Yuzbashyan, B. L. Altshuler, V. B. Kuznetsov, and V. Z. Enolskii, J. Phys. A: Math. Gen. {\bf 38}, 7831 (2005).
\bibitem{bara06} R. A. Barankov and L. S. Levitov, Phys. Rev. Lett. {\bf 96},
  230403 (2006).
\bibitem{alt06} E. A. Yuzbashyan, O. Tsyplyatyev, and B. Altshuler, Phys. Rev. Lett. {\bf 96}, 097005 (2006).
\bibitem{yus06} E. A. Yuzbashyan and M. Dzero, Phys. Rev. Lett. {\bf 96}, 230404 (2006).
\bibitem{randeria14} M. Randeria and E. Taylor, Annu. Rev. Condens. Matter Phys. {\bf 5}, 209 (2014).  

\bibitem{Markiewicz2010b}
R.~S. Markiewicz, J. Lorenzana, G. Seibold, and A. Bansil, Phys. Rev. B {\bf
  81},  014509  (2010).
\bibitem{bueni13} J B\"unemann, M. Capone, J. Lorenzana, and G. Seibold,
  New Journal of Physics {\bf 15}, 053050 (2013).
\bibitem{fabschi1} M. Schir\'{o} and M. Fabrizio, Phys. Rev. Lett. {\bf 105}, 076401 (2010).
\bibitem{fabschi2} M. Schir\'{o} and M. Fabrizio, Phys. Rev. B {\bf 83}, 165105 (2011).  
\bibitem{lor1} G. Seibold and J. Lorenzana, Phys. Rev. Lett. {\bf 86}, 2605 (2001).
\bibitem{seibold03} G. Seibold, F. Becca, and J. Lorenzana, Phys. Rev. B {\bf 67}, 085108 (2003).
\bibitem{seibold04} G. Seibold, F. Becca, P. Rubin, and J. Lorenzana, Phys. Rev. B {\bf 69}, 155113 (2004).  
\bibitem{seibold08} G. Seibold, F. Becca, and J. Lorenzana, Phys. Rev. Lett. {\bf 100}, 016405 (2008); Phys. Rev. B {\bf 78}, 0451124 (2008).
\bibitem{ugenti10} S. Ugenti, M. Cini, G. Seibold, J. Lorenzana, E. Perfetto, and G. Stefanucci, Phys. Rev. B {\bf 82}, 075137 (2010).  
\bibitem{sandri13} M. Sandri and M. Fabrizio, Phys. Rev. B {\bf 88},
  165113 (2013).
\bibitem{mazza17} G. Mazza, Phys. Rev. B {\bf 96}, 205110 (2017).
\bibitem{eckstein09} M. Eckstein, M. Kollar, and P. Werner, Phys. Rev. Lett. {\bf 103}, 056403 (2009). 
\bibitem{werner12} P. Werner, N. Tsuji, and M. Eckstein, Phys. Rev. B {\bf 86}, 205101 (2012).
\bibitem{tsuji13} N. Tsuji, M. Eckstein, and P. Werner, Phys. Rev. Lett. {\bf 110}, 136404 (2013).  
\bibitem{balzer15} K. Balzer, F. A. Wolf, I. P. McCulloch, P. Werner, and
  M. Eckstein, Phys. Rev. X {\bf 5}, 031039 (2015).
\bibitem{sandri2} M. Sandri, M. Schir\'{o}, and M. Fabrizio, Phys. Rev. B {\bf 86}, 075122 (2012).  
\bibitem{mazza12} G. Mazza and M. Fabrizio, Phys. Rev. B {\bf 86}, 184303 (2012).
\bibitem{Micnas1990}
R. Micnas, J. Ranninger, and S. Robaszkiewicz, Rev. Mod. Phys. {\bf 62},  113
  (1990).

\bibitem{Medina1991}
G.~A. Medina, J. Simonin, and M.~D. {N{\'{u}}{\~{n}}ez Regueiro}, Phys. Rev. B
  {\bf 43},  6206  (1991).

\bibitem{Sofo1992}
J.~O. Sofo and C.~A. Balseiro, Phys. Rev. B {\bf 45},  377  (1992).

\bibitem{goetz108} G. Seibold, F. Becca, and J. Lorenzana
Phys. Rev. Lett. {\bf 100}, 016405 (2008).   
\bibitem{goetz208} G. Seibold, F. Becca, and J. Lorenzana,
  Phys. Rev. B {\bf 78}, 045114 (2008).
\bibitem{br}  W. F. Brinkman and T. M. Rice, Phys. Rev. B {\bf 2}, 4302 (1970).
\bibitem{volk74} A. F. Volkov and Sh. M. Kogan, Zh. Eksp. Teor. Fiz. {\bf 65}, 2038 (1973)
  [Sov- Phys. JETP {\bf 38}, 1018 (1974].
\bibitem{slichter} C. P. Slichter, Principles of Magnetic Resonance, Springer
Series in Solid-State Sciences Vol. 1 (Springer, Berlin, 1990),
p. 655.
\bibitem{tohyama} C. Shao, T. Tohyama, H.-G. Luo, and H. Lu, Phys. Rev. B {\bf 93}, 195144 (2016).
\bibitem{comment} More precisely the limit $\omega\to 0$ has to be taken before
  the limit $q\to 0$ for a momentum and frequency dependent vector potential
  $A(q,\omega)$.
\end{thebibliography}
\end{document}